\documentclass[namedreferences]{solarphysics}
\usepackage[hyperref,optionalrh,solaromanenum]{spr-sola-addons} 
\usepackage{graphicx}                    
\usepackage{color}                       
\usepackage{breakurl}                         

\providecommand{\e}[1]{\ensuremath{\times 10^{#1}}}

\newcommand{\aap}{    {\it Astron. Astrophys.}}

\newcommand{\araa}{ {\it Annu. Rev. Astron. Astrophys.}}

\newcommand{\apj}{    {\it Astrophys. J.}}
\newcommand{\apjl}{   {\it Astrophys. J. Lett.}}

\newcommand{\gafd}{   {\it Geophys. Astrophys. Fluid Dyn.}}

\newcommand{\mnras}{  {\it Mon. Not. Roy. Astron. Soc.}}
\newcommand{\nat}{    {\it Nature}}

\newcommand{\solphys}{{\it Solar Phys.}}

\newcommand{\lrsp}{    {\it Living Rev. Solar Phys.}}

\begin{document}

\begin{article}

\begin{opening}

\title{Effects of Radiative Diffusion on Thin Flux Tubes in Turbulent Solar-like Convection}

%
\author{M.A.~\surname{Weber}$^{1}$\sep
        Y.~\surname{Fan}$^{2}$
       }

%
\runningauthor{M.A. Weber, Y. Fan}
\runningtitle{Effects of Radiative Diffusion and Convection on Rising Flux Tubes}

%
  \institute{$^{1}$ Department of Physics and Astronomy, University of Exeter, Stocker Road, Exeter, UK EX4 4QL
                     email: \url{mweber@astro.ex.ac.uk} \\ 
                $^{2}$ High Altitude Observatory, National Center for Atmospheric Research, 3080 Center Green Drive, Boulder, CO, USA 80301
                     email: \url{yfan@ucar.edu} \\
             }

\begin{abstract}
We study the combined effects of convection and radiative diffusion on the evolution of thin magnetic flux tubes in the solar interior.  Radiative diffusion is the primary supplier of heat to convective motions in the lower convection zone, and it results in a heat input per unit volume of magnetic flux tubes that has been ignored by many previous thin flux tube studies.  We use a thin flux tube model subject to convection taken from a rotating spherical shell of turbulent, solar-like convection as described by Weber, Fan, and Miesch (2011, \textit{Astrophys. J.}, \textbf{741}, 11; 2013, \textit{Solar Phys.}, \textbf{287}, 239), now taking into account the influence of radiative heating on  $10^{22}$ Mx flux tubes, corresponding to flux tubes of large active regions.  Our simulations show that flux tubes of $\le$60 kG subject to solar-like convective flows do not anchor in the overshoot region, but rather drift upward due to the increased buoyancy of the flux tube earlier in its evolution as a result of the inclusion of radiative diffusion.  Flux tubes of magnetic field strengths ranging from 15 kG to 100 kG have rise times of $\le$0.2 years, and exhibit a Joy's Law tilt-angle trend.  Our results suggest that radiative heating is an effective mechanism by which flux tubes can escape from the stably stratified overshoot region, and that flux tubes do not necessarily need to be anchored in the overshoot region to produce emergence properties similar to those of active regions on the Sun.
\end{abstract}

%

\end{opening}

%

\section{Introduction}
\label{sec:intro} 
Sunspots are observable manifestations of magnetism on the solar surface, thus providing a link to the deep-seated dynamo mechanism.  For the Sun, the prevailing dynamo paradigm suggests that the toroidal magnetic field giving rise to the longitudinal bands of solar activity is amplified and stored in a thin shearing region at the base of the convection zone called the tachocline \citep[\textit{e.g.}][]{gilman_2000,char_lrsp_2010}.  Buoyant loops of magnetic flux must then traverse the bulk of the turbulent convection zone, eventually emerging at the surface to form active regions.

The thin flux tube (TFT) approximation has been developed by a number of authors to describe the dynamics of thin, isolated magnetic flux tubes \citep[\textit{e.g.}][]{defouw_1976,roberts_1978,spruit_1981a,spruit_1981b,moreno_1986,ferriz_1989,cheng_1992,ferriz_1993,achterberg_1996,moreno_1996}.  Numerical simulations based on the TFT approach have provided valuable insight on the evolution of rising magnetic loops through a quiescent solar convection zone \citep[see review by][]{fan_2009}.  More recently, \citet{weber_2011} (hereafter Article 1) and \citet{weber_solphys_2013} (hereafter Article 2) have extended the TFT model to include the effects of a rotating spherical shell of turbulent solar-like convective flows on the evolution of rising flux tubes.  Most previous TFT studies assume that flux tubes evolve adiabatically throughout the convection zone (exceptions to this are \textit{e.g.} \citet{fan_1996,moreno_2002,rempel_2003}).  This is a valid assumption for the upper $\approx$2/3 of the solar convection zone. However, in the lower $\approx$1/3 of the convection zone, closer to the radiative interior of the Sun, there is a significant non-zero divergence of radiative heat flux due to the deviation from radiative equilibrium.  Radiative diffusion in the lower convection zone is the primary supplier of heat to large-scale global convective motions.  It also results in a heat input per unit volume [$Q_{v}$] of magnetic flux tubes, which may have a substantial effect on their buoyancy, and hence their dynamic evolution.

Due to radiative heating, \citet{fan_1996} found that flux tubes of $10^{21}-10^{22}$ Mx rise through the convection zone in $\approx$two to four months, a comparable or shorter timescale compared to flux tubes allowed to evolve adiabatically, which have rise times of $\approx$two to ten months \citep{fan_1993}.  This shorter rise time is a result of radiative heating near the bottom of the convection zone, which increases the density deficit (\textit{i.e.} buoyancy) of the flux tube at deeper depths, initiating more quickly their buoyant rise toward the surface.  These emerging flux loops also show similar qualitative features to solar active regions such as tilt angles and morphological asymmetries, not significantly different from simulation results where flux tubes evolve adiabatically \citep[\textit{e.g.}][]{fan_1993,dsilva_1993, caligari_1995,caligari_1998}.  \citet{fan_1996} also find that flux tubes subject to radiative diffusion will rise quasi-statically (\textit{i.e.} all forces closely balance) through the convective overshoot region.  Using typical values for the overshoot region as computed by numerical solar-structure models, flux tubes take $\approx$one year or less to emerge from the overshoot region \citep[\textit{e.g.}][]{vanballe_1982,fan_1996,rempel_2003}.  This short storage time compared to the $\approx$11-year solar cycle may have significant implications for the solar dynamo mechanism. 

The purpose of this article is to study the combined effects of turbulent solar-like convection and heating due to radiative diffusion on the evolution of active-region-scale $10^{22}$ Mx magnetic flux tubes. We begin in Section \ref{sec:model} with a description of our simulation model and the modification to the TFT energy equation due to the addition of the non-adiabatic heating by radiative diffusion.  In Section \ref{sec:dynamics_rad} we address how the addition of radiative diffusion to the energy equation alters the dynamic evolution of flux tubes subject to solar-like convective flows.  Emergence properties of these flux tubes, specifically latitude of emergence and tilt-angle properties, will be compared in Sections \ref{sec:latem_rad} and \ref{sec:tilt_rad} respectively, to those of flux tubes that evolve adiabatically.  We will mention the problem of flux storage in Section \ref{sec:flux_storage}, and end with a summary and discussion in Section \ref{sec:discuss_rad}.                       
  
\section{Model Description}
\label{sec:model}

\subsection{The TFT+ASH Approach}
\label{sec:hybrid}

We model the evolution of magnetic flux tubes in the solar interior following the thin flux tube (TFT) approximation.  The TFT equations are derived from the ideal MHD equations, operating under the assumption that the flux tube radius [$a$] is small compared to the total length [$L$], and other relevant length-scale variations such as the local pressure scale height and radius of curvature of the flux tube.  All physical quantities of the flux tube are taken as averages over the cross-section, only varying spatially along the flux tube axis.  The TFT equations that describe the evolution of each Lagrangian element of the one-dimensional flux tube are as follows (and similar for Articles 1 and 2):
\begin{eqnarray}
\rho {\mathrm{d} {\mathbfit{v}} \over \mathrm{d}t} & = & -2 \rho ( {\mathbfit{\Omega}_0} \times {\mathbfit{v}} )
-(\rho_\mathrm{e} - \rho ) [{\mathbfit{g}} - {\mathbfit{\Omega}_\mathrm{0}} \times ({\mathbfit{\Omega}_\mathrm{0}} \times
{\mathbfit{r}})] + {\mathbfit{l}} {\partial \over \partial s} \left ( { B^2 \over 8 \pi}
\right ) + {B^2 \over 4 \pi} {\mathbfit{k}} 
\nonumber \\
& & - C_\mathrm{d} {\rho_\mathrm{e} | ({\mathbfit{v} }-{\mathbfit{v}}_\mathrm{e})_{\mathrm{\perp}} |
({\mathbfit{v}}-{\mathbfit{v}}_\mathrm{e})_{\mathrm{\perp}} \over ( \pi \Phi / B )^{1/2} }, 
\label{eq:eqn_motion_chmodel}
\end{eqnarray}
\begin{equation}
{\mathrm{d} \over \mathrm{d}t} \left ( {B \over \rho} \right )  =  {B \over \rho} \left [
{\partial ({\mathbfit{v}} \cdot {\mathbfit{l}}) \over \partial s} - {\mathbfit{v}} \cdot
{\mathbfit{k}} \right ] , 
\label{eqn_cont_induc}
\end{equation}
\begin{equation}
{1 \over \rho} {\mathrm{d} \rho \over \mathrm{d}t}  =  {1 \over \gamma p} {\mathrm{d}p \over \mathrm{d}t} - {\nabla_{\mathrm{ad}}} {\frac{\rho}{p} T \frac{\mathrm{d}S}{\mathrm{d}t}} ,
\label{eq:eqn_energy}
\end{equation}
\begin{equation}
p  =  {\rho R T \over \mu} ,
\label{eqn_state}
\end{equation}
\begin{equation}
p + {B^2 \over 8 \pi}  =  p_\mathrm{e} ,
\label{eq:eqn_pbalance}
\end{equation} 
where, ${\mathbfit{r}}$, ${\mathbfit{v}}$, $B$, $\rho$, $p$, $T$, which are functions of the time [$t$] and arc length [$s$] measured along the tube, denote respectively the position, velocity, magnetic field strength, gas density, pressure, and temperature of a Lagrangian tube segment, ${\mathbfit{l}} \equiv \partial {\mathbfit{r}} / \partial s$ is the unit vector tangential to the flux tube, ${\mathbfit{k}} \equiv \partial^2 {\mathbfit{r}} / \partial s^2 $ is the tube's curvature vector, subscript ${\perp}$ denotes the component perpendicular to the flux tube, $\Phi$ is the constant total flux of the tube, $\rho_e$, $p_e$, and $\mu$, which are functions of depth only, are respectively the pressure, density, and mean molecular weight of the surrounding external plasma, ${\mathbfit{g}}$ is the gravitational acceleration that is a function of depth only, ${\mathbfit{\Omega}_\mathrm{0}}$ is the angular velocity of the reference frame co-rotating with the Sun, with $\Omega_\mathrm{0}=2.7 \times 10^{-6}$ rad s$^{-1}$, $C_\mathrm{d}=1$ is the drag coefficient, $\gamma$ is the ratio of specific heats, $S$ is the entropy per unit mass, $\nabla_\mathrm{ad}$ is the adiabatic temperature gradient, and ${\mathbfit{v}}_\mathrm{e} ({\mathbfit{r}}, t)$ is a time dependent velocity field (relative to the rotating frame of reference) that impacts the dynamics of the thin flux tube through the drag force term (last term in Equation (\ref{eq:eqn_motion_chmodel})).  The term ${\mathbfit{v}}_\mathrm{e}$ accounts for both the local convective flows and mean flows such as differential rotation.  Heating due to radiative diffusion is introduced through the last term on the right-hand side of Equation (\ref{eq:eqn_energy}), which we will describe in detail in Section \ref{sec:energy}. 

In the above equations, we do not introduce an explicit magnetic-diffusion or kinematic-viscosity term.   The thin flux tube is untwisted (\textit{i.e.} magnetic field lines do not twist about the flux tube axis), and is discretized with 800 uniformly spaced grid points along its arc length [$s$].  The numerical methods used to solve for the flux tube evolution as determined by the above set of equations have been described in detail by \citet{fan_1993}.  

For the stratification and thermodynamic properties of the external field-free plasma, namely $\rho_\mathrm{e}$, $p_\mathrm{e}$, $T_\mathrm{e}$, $g$, and $\delta=\nabla_\mathrm{e}-\nabla_\mathrm{ad}$, where $\nabla_\mathrm{e}=\mathrm{d} \ln{T_\mathrm{e}}/\mathrm{d} \ln{\rho_\mathrm{e}}$ and $\nabla_\mathrm{ad}$ is the value of $\nabla_\mathrm{e}$ that one obtains by considering local adiabatic perturbations, we use the reference solar model (hereafter Model S) by \citet{jcd_1996}.  See Figure 1 in Article 1 for profiles of $T_\mathrm{e}$, $\rho_\mathrm{e}$, $p_\mathrm{e}$, and $\delta$ used in our simulations.  Here we define the base of the convection zone as $r_\mathrm{czb}=$ 0.723\,R$_{\odot}$ (5.026\e{10} cm), the radius in Model S where the plasma changes from sub-adiabatic (stably stratified, $\nabla_\mathrm{e}<\nabla_\mathrm{ad}$) to super-adiabatic (unstably stratified, $\nabla_\mathrm{e}>\nabla_\mathrm{ad}$).  Model S contains no region of convective penetration below $r_\mathrm{czb}$.  To mimic a convective overshoot region, we extend Model S below $r_\mathrm{czb}$ with a simple polytropic, sub-adiabatically stratified region that decreases from values of $\delta=0$ to $-10^{-3}$ over a distance of $\approx0.01$\,R$_{\odot}$ ($\approx10^{9}$ cm) downward from $r_\mathrm{czb}$. In this region, it is assumed that overshooting convective motions are efficient enough to establish a sub-adiabatic stratification such that the temperature gradient remains closer to the adiabatic value rather than the radiative temperature gradient of the deep interior (see Section \ref{sec:solar_overshoot} for more details).

What sets the thin flux tube simulations introduced in Articles 1 and 2, and continued here, apart from previous simulations is the inclusion into the drag force term of a time-dependent convective velocity field ${\mathbfit{v}}_\mathrm{e} ({\mathbfit{r}}, t)$ relative to the rotating frame of reference.  This three-dimensional global convection simulation is computed separately using the Anelastic Spherical Harmonic (ASH) code, as described by \citet{miesch_apj_2006}.  The ASH code solves the 3D anelastic Navier--Stokes fluid equations using a pseudo-spectral method with both spherical harmonic and Chebyshev basis functions, explicitly resolving the largest scales of motion, while treating small turbulent eddies with sub-grid techniques.

The computational domain for the ASH simulation extends from $r = 0.69$\,R$_{\odot}$ to $r = 0.97$\,R$_{\odot}$ (4.8\e{10} cm to 6.75\e{10} cm), and the density contrast across the domain is $\approx$69, or 4.2 density scale heights.  The Rayleigh number [$R_\mathrm{a} = g r^2 d \Delta S / (\nu \kappa C_\mathrm{P})$] is 5\e{6} in the mid-convection zone, where $d = r_2-r_1$ is the depth of the domain, $C_\mathrm{P}$ is the specific heat at constant pressure, and $\nu$ and $\kappa$ are respectively the turbulent viscosity and thermal diffusivity of the ASH simulation.  The Reynolds number [$R_\mathrm{e} = v_\mathrm{rms} d / \nu$] is of order 50, where $v_\mathrm{rms}$ is the root-mean-square velocity relative to the rotating reference frame.  For more specific details on the ASH simulation used here, see Articles 1 and 2.

\begin{figure} 
\centerline{
\includegraphics[scale=.72, angle=90]{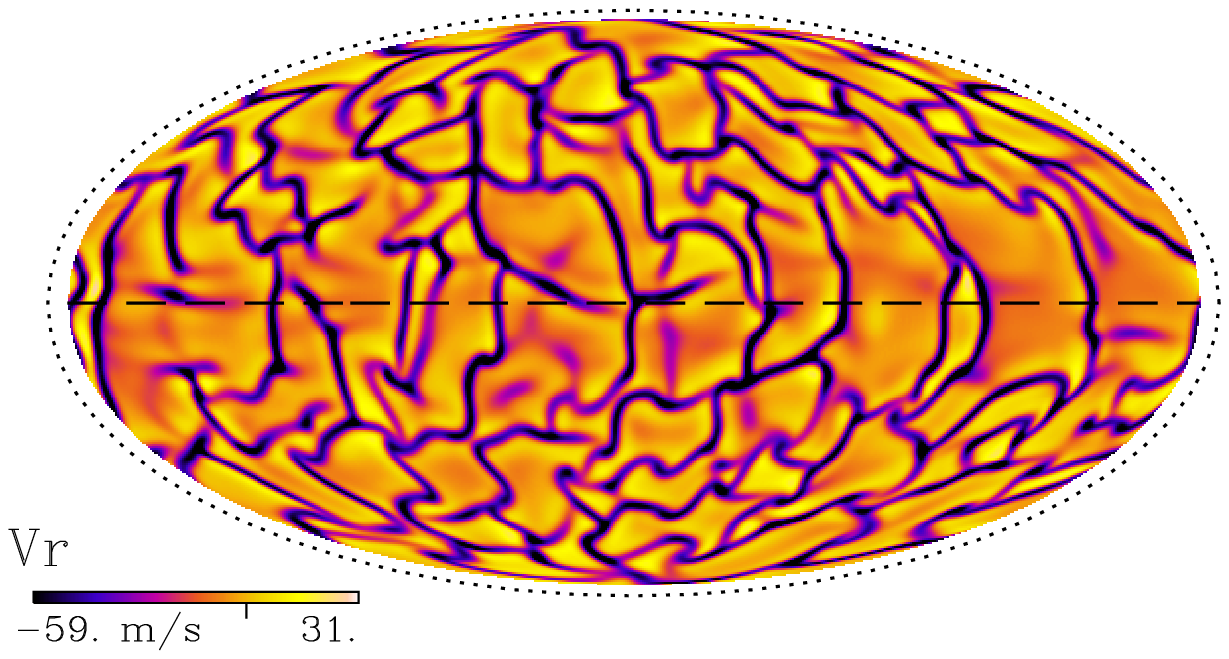}
\includegraphics[trim=0cm 10.9cm 0cm 0cm,clip=true,scale=.33,angle=90]{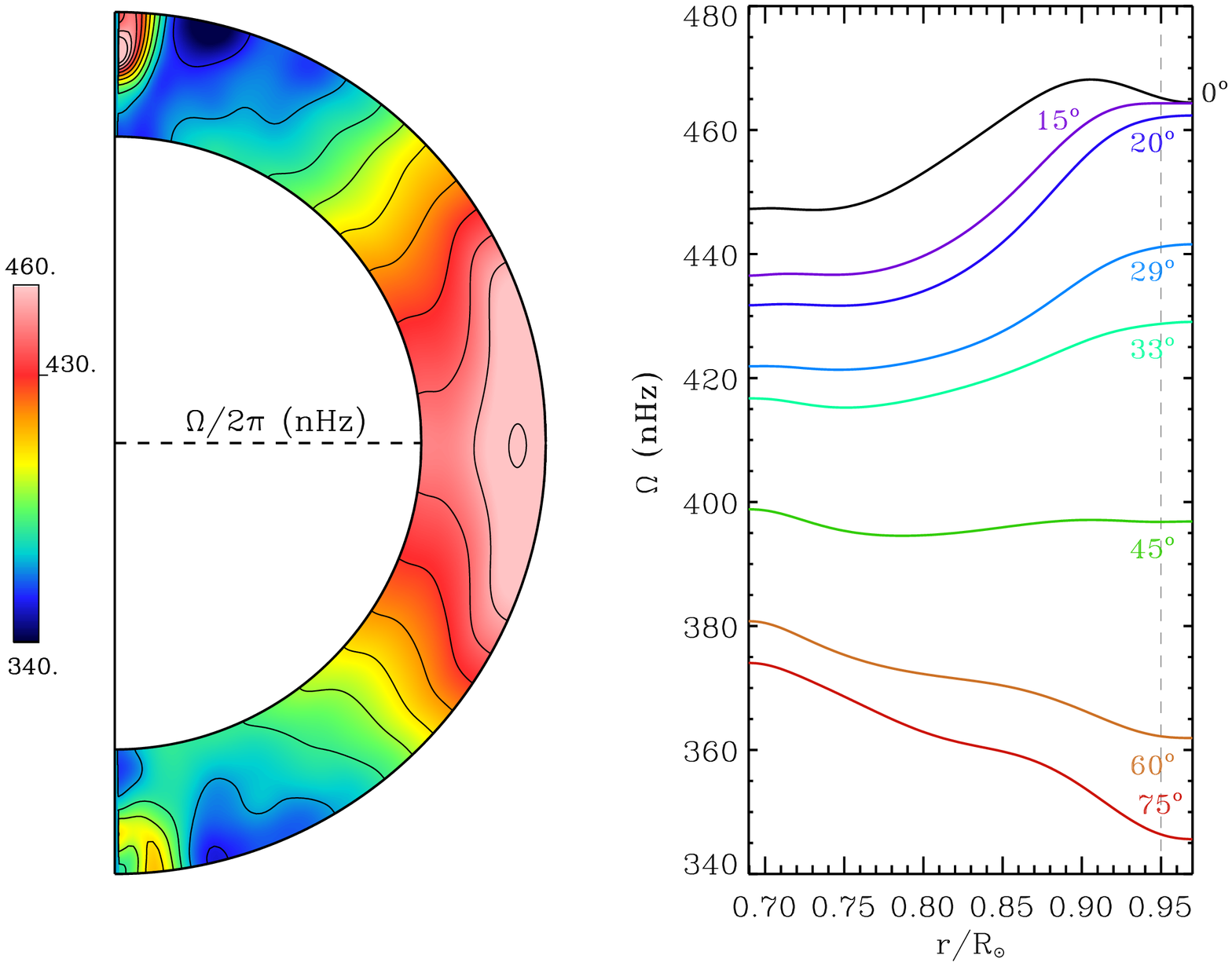}}
\caption{(Left) Snapshot of ASH simulation convective radial velocities used in this article at a depth of 23 Mm in a Mollweide projection, with the dotted line representing the solar radius $r=$ R$_{\odot}$.  Strong downflow lanes (purple) at the boundary of giant convective cells surround upflow regions (yellow).  Also known as banana cells, the structures at low latitudes are rotationally aligned and propagate prograde. (Right) Angular velocity (with respect to the inertial frame), averaged over longitude and time (time interval 755 days).  Color table saturates at the values indicated, with extrema ranging from 326 nHz to 468 nHz.}
\label{fig:radial_velocities}
\end{figure}

A typical giant-cell convection pattern of this ASH simulation at a depth of 23 Mm below the solar surface is shown on the left in Figure \ref{fig:radial_velocities}. Broad upflow cells are surrounded by narrow downflow lanes, which can reach maximum downflow speeds of nearly $600$ m s$^{-1}$ at a mid-convection zone depth of about $86$ Mm below the surface.  These downflow lanes align preferentially with the rotation axis at low latitudes, reflecting the presence of so-called ``banana cells''.  The total angular velocity of the convection simulation $\Omega/2 \pi$ (with respect to the inertial frame) is solar-like, and decreases monotonically from $\approx$470 nHz at the Equator to $\approx$330 nHz at the Poles, and exhibits nearly conical contours at mid-latitudes (see Figure \ref{fig:radial_velocities}, right), as observed in the solar convection zone \citep[\textit{e.g.}][]{thompson_araa_2003}.  The combined influence of density stratification and the Coriolis force induces anti-cyclonic vorticity in expanding upflows and cyclonic vorticity in contracting downflows.  These effects yield a mean kinetic helicity density [$H_\mathrm{k}$] that is negative in the northern hemisphere and positive in the southern hemisphere (see Article 1 for an image of the associated kinetic helicity).   Such a helicity pattern is typical for rotating, compressible convection \citep[\textit{e.g.}][]{miesch_2009}.

We emphasize that we use the one-dimensional Model S to describe the background stratification and thermodynamic properties of the computation domain, utilizing the ASH simulation only for the convective flow field.  This choice was made to facilitate comparison between flux tubes allowed to evolve both with and without convection (see Section \ref{sec:dynamics_rad}).  Differences in results will then be solely due to the presence of convective flows or the form of the TFT energy equation (see Section \ref{sec:energy}).  In any case, the ASH background reference state is not significantly different from Model S, by at most $\approx$$10\%$ throughout the majority of the simulation domain.  Time-varying temperature fluctuations in the ASH simulation from its prescribed reference state are small, on the order of $\delta$ in the bulk of the convection zone.  Furthermore, the ASH simulation is computed in the hydrodynamic regime separately from the TFT simulations.  Any thermodynamic changes associated with the presence of the magnetic field in a truly three-dimensional magnetohydrodynamic ASH simulation would be on the order of $1/\beta$, where $\beta=8\pi p_\mathrm{e}/B^2$.

Our simulations start with toroidal magnetic flux rings in mechanical equilibrium located at a radial distance to the center of the Sun $r = r_{\rm 0} = 5.05 \times 10^{10}$ cm, slightly above the base of the solar convection zone at $r = r_{\rm czb} = 5.026 \times 10^{10}$ cm.  At $r_\mathrm{0}$, the super-adiabaticity is $\delta\approx$ 3\e{-8}.  To ensure the initial state of mechanical equilibrium, the flux tube is made neutrally buoyant by reducing the internal temperature compared to the external temperature and the magnetic tension force is balanced by the Coriolis force from an initially prograde toroidal flow in the tube.  The external time-dependent convective velocity field described impacts the flux tube through its associated drag force (last term in Equation (\ref{eq:eqn_motion_chmodel})).  

We consider a range of flux tubes with initial magnetic field strengths [$B_\mathrm{0}$] of 15, 30, 40, 50, 60, 80, and 100 kG, and initial latitudes [$\theta_\mathrm{0}$] ranging from $1^{\circ} $ to $40^{\circ}$ in either hemisphere.  Considering that the root-mean-square (rms) of the convective downflows from the ASH simulation at the base of the convection zone is $\approx$35 m s$^{-1}$, the equipartition magnetic field strength for our simulation is the $B_\mathrm{eq}\approx5$ kG.  In this case, we are investigating flux tubes on the order of $3-20$ $B_\mathrm{eq}$.  The flux of the tube is constant at $10^{22}$ Mx for this article, typical of large active regions with the strongest sunspots.  As was also done in Articles 1 and 2, each flux tube is initially perturbed with small undular motions consisting of a superposition of Fourier modes with azimuthal order ranging from $m=0$ to $m=8$ with random phase relations.  However, such perturbations have negligible effects on the evolution of flux tubes subject to convection because perturbations provided by the convective velocity field are much stronger in amplitude. 

We perform seven groups of simulations sampling different time ranges of the ASH convective flow at each magnetic field strength and initial latitude in both the northern and southern hemispheres, for a total number of $\approx$2300 individual flux tube simulations. The flux tubes comprising one group are released at the base of the convection zone at the same starting time, although they do not interact with each other (\textit{i.e.} are isolated) and are allowed to evolve until some portion of the flux tube reaches the top of the simulation domain.  The flux tube release times for the groups are arbitrarily chosen, but are at least separated by the convective turnover timescale of the ASH convection simulation, which is $\approx$30 days.  In this way, the flux tubes are able to sample significantly different portions of the convective velocity field.  This procedure is the same as was done in Article 2.

\subsection{The Thin Flux Tube Energy Equation}
\label{sec:energy}

In the radiative interior of the Sun, energy is transported only by radiation.  However, in the outer third of the Sun by radius, convection begins to take over much of the energy transport.  Therefore, in the convection zone and overshoot region, the plasma is no longer in radiative equilibrium, resulting in a non-zero divergence of radiative heat flux.  This acts to heat the flux tube in the lower convection zone, and cool it near the surface.  Additionally, there is radiative diffusion across the flux tube due to a difference in temperature between the interior of the flux tube and the surrounding plasma \citep[see, \textit{e.g.},][]{parker_book,vanballe_1982}.  As a result, especially in the lower convection zone, it does not suffice to assume that the flux tube evolves adiabatically.  The incorporation of radiative heating to the TFT equations has been described in detail by \citet{fan_1996}.  Here we briefly review their methods for clarity in this article. 

The rate of heat input per unit volume [d$Q_\mathrm{v}$/dt] of the flux tube plasma is \citep{fan_1996}:
\begin{equation}
\rho T \frac{\mathrm{d}S}{\mathrm{d}t} =  \frac{\mathrm{d}Q_\mathrm{v}}{\mathrm{d}t} = \nabla \cdot (\kappa \nabla T),
\label{eq:eqn_dqdt}
\end{equation}
where $S$ is the entropy per unit mass, the right-most expression is a divergence of a radiative diffusive heat flux [$\kappa \nabla T$], and $\kappa$ is the coefficient of radiative conductivity.  Note that in the adiabatic regime d$S/$d$t=0$.  This is the assumption made in Articles 1 and 2.

Following \citet{fan_1996}, Equation (\ref{eq:eqn_dqdt}) can be reduced to two dominant terms:
\begin{equation}
\frac{\mathrm{d}Q_{v}}{\mathrm{d}t} \approx \nabla \cdot (\kappa_\mathrm{e} \nabla T_\mathrm{e}) + \kappa \nabla_{\perp}^{2} \delta T,
\label{eq:dqdt_reduced}
\end{equation} 
where $\kappa_\mathrm{e}$ and $\kappa$ are the coefficient of radiative conductivity of the background field-free plasma and the flux tube plasma, respectively.  The term $\delta T = T-T_\mathrm{e}$ represents a small difference between the external and internal temperature introduced due to the presence of a magnetic field in the flux tube, and $\nabla_{\perp}$ is the gradient vector in the direction perpendicular to the flux tube axis (in the plane of the flux tube cross-section).  The first term on the right-hand side of Equation (\ref{eq:dqdt_reduced}), which we will subsequently refer to as $(\mathrm{d}Q/\mathrm{d}t)_{1}$, is dependent only on the external plasma thermodynamic quantities $\kappa_\mathrm{e}$ and $T_\mathrm{e}$, which vary only as a function of radial distance from Sun center.  This term is zero if the fluid is in radiative equilibrium, which is the case in the radiative interior of the Sun.  The second term, subsequently referred to as $(\mathrm{d}Q/\mathrm{d}t)_{2}$, represents a radiative diffusion across the flux tube due to the temperature difference between the flux tube and the external plasma.  The two heating terms are computed as follows \citep{fan_1996}:
\begin{equation}
\biggl( \frac{\mathrm{d}Q}{\mathrm{d}t} \biggr)_{1} = -F_\mathrm{tot} \frac{\mathrm{d}}{\mathrm{d}r} \biggl( \frac{\nabla_\mathrm{e}}{\nabla_\mathrm{rad}}  \biggr), 
\label{eqn:dq_dt1_final}
\end{equation} 
\begin{equation}
\biggl( \frac{\mathrm{d}Q}{\mathrm{d}t}  \biggr)_{2} \approx -\kappa_\mathrm{e} \frac{\alpha_{1}^{2}}{a^{2}} ({T}-{T_\mathrm{e}}),
\end{equation}
where the magnitude of the total energy flux is given by $F_\mathrm{tot}=L/(4\pi r^{2})$, and $L$ is the total luminosity of the Sun, $\nabla_\mathrm{e}= \mathrm{d} \ln{T_\mathrm{e}}/\mathrm{d} \ln{p_\mathrm{e}}$ is the local temperature gradient of the external plasma environment, and $\nabla_\mathrm{rad}=(\mathrm{d} \ln{T_\mathrm{e}}/\mathrm{d} \ln{p_\mathrm{e}})_\mathrm{rad}$ is the temperature gradient required for the energy to be transported only by radiation, $\alpha_{1} \approx 2.4048$ is the first zero of the Bessel function $J_{0}(x)$, $a$ is the flux tube radius, and ${\delta T}={T}-{T_\mathrm{e}}$ is the mean temperature difference across the flux tube.              
       
The magnitude of $(\mathrm{d}Q/\mathrm{d}t)_{1}\approx\kappa_\mathrm{e}\nabla_\mathrm{e}T_\mathrm{e}/H_\mathrm{p}^2$, where $H_\mathrm{p}=p_\mathrm{e}/g \rho_\mathrm{e}$ is the pressure scale height and we have used $\mathrm{d}/\mathrm{d}r=1/H_\mathrm{p}$ and $F_\mathrm{tot}=\kappa_\mathrm{e}T_\mathrm{e}\nabla_\mathrm{rad}/H_\mathrm{p}$ \citep{fan_1996}.  For a neutrally buoyant flux tube at the base of the convection zone $\rho=\rho_\mathrm{e}$.  Such that the condition of pressure balance (Equation (\ref{eq:eqn_pbalance})) is satisfied, the internal temperature of the flux tube must be less than the surrounding environment.  This temperature deficit $(\delta T/T_\mathrm{e})\approx 1/\beta = B^{2}/8\pi p_\mathrm{e}$, where $\beta \gg 1$ in the solar interior.  Then, comparing the magnitudes of $(\mathrm{d}Q/\mathrm{d}t)_{1}$ and $(\mathrm{d}Q/\mathrm{d}t)_{2}$ to determine their relative importance to the energy equation \citep{fan_1996}:
\begin{equation}
\frac{\biggl | \biggl( \frac{\mathrm{d}Q}{\mathrm{d}t} \biggr)_{1}   \biggr | }{\biggl | \biggl( \frac{\mathrm{d}Q}{\mathrm{d}t} \biggr)_{2}   \biggr | } \approx  \frac{\nabla_\mathrm{e}}{\alpha_{1}^{2}} \frac{a^{2}}{H_\mathrm{p}^{2}} \frac{T_\mathrm{e}}{\delta T} \approx \frac{8 \nabla_\mathrm{e}}{\alpha_{1}^{2}} \frac{p_\mathrm{e}}{H_\mathrm{p}^{2}} \frac{\Phi}{B^{3}}.
\label{eqn:ratio}
\end{equation}        
Throughout the convection zone $\nabla_\mathrm{e} \approx 0.4$.  Taken from Model S, typical values of the pressure scale height and pressure at $r_\mathrm{0}=5.05\e{10}$ cm near the base of the convection zone are $H_\mathrm{p}\approx$ 5.5\e{9} cm and $p_\mathrm{e}\approx$ 4.7\e{13} g s$^{-2}$ cm$^{-1}$.  Therefore, the magnitude of Equation (\ref{eqn:ratio}) is $\approx$8.6\e{-7} $\Phi/B^{3}$.  For $\Phi=10^{22}$ Mx, the ratios of $|(\mathrm{d}Q/\mathrm{d}t)_{1}|$ to $|(\mathrm{d}Q/\mathrm{d}t)_{2}|$ following this relationship are given in Table \ref{tbl:ratio}.  As Equation (\ref{eqn:ratio}) is proportional to $\Phi$, the values quoted in Table \ref{tbl:ratio} will decrease by an order of magnitude for $\Phi=10^{21}$ Mx, and by two orders of magnitude for $\Phi=10^{20}$ Mx.        

The second heating term [$(\mathrm{d}Q/\mathrm{d}t)_{2}$] acts to reduce the temperature difference between the flux tube and the external plasma environment, bringing it closer to a state of thermal equilibrium, enhancing the density deficit of the flux tube in the process.  This term is inversely proportional to the flux tube radius, therefore it will increase in magnitude for thinner tubes (\textit{i.e.} as the magnetic field strength increases and magnetic flux decreases).  However, the first heating term [$(\mathrm{d}Q/\mathrm{d}t)_{1}$] is dependent only on properties of the background plasma as a function of distance [$r$] from Sun center.  Therefore each flux tube will experience the same heating from the term $(\mathrm{d}Q/\mathrm{d}t)_{1}$ at the same distance [$r$].  We have chosen in this article to focus on how radiative heating in conjunction with convection influences the dynamic properties of $10^{22}$ Mx flux tubes, corresponding to large solar active regions.  Following the values reported in Table \ref{tbl:ratio}, it is a reasonable approximation to neglect the term $(\mathrm{d}Q/\mathrm{d}t)_{2}$ in the TFT energy equation for flux tubes of $10^{22}$ Mx in the 15\,--\,100 kG range, as was done by \citet{fan_1996}.  Future TFT studies incorporating the effects of radiative heating on flux tubes of $<10^{22}$ Mx will require the inclusion of $(\mathrm{d}Q/\mathrm{d}t)_{2}$, especially at larger magnetic field strengths. Combining the general TFT energy equation (Equation (\ref{eq:eqn_energy})) with Equation (\ref{eqn:dq_dt1_final}), the new energy equation for the TFT model we use is \citep{fan_1996}:
\begin{eqnarray}
{1 \over \rho} {\mathrm{d} \rho \over \mathrm{d}t} &=& {1 \over \gamma p} {\mathrm{d}p \over \mathrm{d}t} + {\nabla_\mathrm{ad}} {\frac{F_\mathrm{tot}}{p} \frac{\mathrm{d}}{\mathrm{d}r} \biggl(\frac{\nabla_\mathrm{e}}{\nabla_\mathrm{rad}} \biggr)},
\label{eqn:energy_rad}
\end{eqnarray}           
where the last term was assumed to be zero for the adiabatically evolving TFT simulations of Articles 1 and 2.  Values for the quantities $\nabla_\mathrm{ad}$, $\nabla_\mathrm{e}$, $\nabla_\mathrm{rad}$, and $F_\mathrm{tot}$ are provided by Model S.    

\begin{table}
\begin{tabular}{cccccccc}
\\
\hline
 & 15 kG & 30 kG & 40 kG & 50 kG & 60 kG & 80 kG & 100 kG \\
\hline
$|\mathrm{d}Q_{1}/\mathrm{d}Q_{2}|$ &  2500 & 320 & 130 & 69 & 40 & 17 & 8.6 \\
\hline
\end{tabular}
\caption{Ratio of $|(\mathrm{d}Q/\mathrm{d}t)_{1}|$ to $|(\mathrm{d}Q/\mathrm{d}t)_{2}|$ following Equation (\ref{eqn:ratio}) for flux tubes of $10^{22}$ Mx.  The term $|(\mathrm{d}Q/\mathrm{d}t)_{1}|$ is much larger than $|(\mathrm{d}Q/\mathrm{d}t)_{2}|$ for tubes of $10^{22}$ Mx.  As $|(\mathrm{d}Q/\mathrm{d}t)_{1}|/|(\mathrm{d}Q/\mathrm{d}t)_{2}|$ is directly proportional to $\Phi$, the quoted values will decrease by an order of magnitude for $\Phi=10^{21}$ Mx, and by two orders of magnitude for $\Phi=10^{20}$ Mx.  Values have been rounded to two significant digits.}
\label{tbl:ratio}
\end{table} 

This added heating term (last term of Equation (\ref{eqn:energy_rad})) affects the dynamic evolution of the flux tube by changing its buoyancy evolution [$\mathrm{d}\Delta \rho/\mathrm{d}t$].  To illustrate the importance of including the radiative-heating term in the TFT energy equation, \citet{fan_1996} examined the growth of buoyancy caused by radiative heating as compared to that due to the adiabatic expansion of the flux tube plasma rising through a super-adiabatically stratified environment:
\begin{equation}
\biggl( \frac{\mathrm{d} \Delta \rho}{\mathrm{d}t} \biggr)_\mathrm{rad} = \frac{\rho_\mathrm{e}}{p_\mathrm{e}}\nabla_\mathrm{ad} \biggl( \frac{\mathrm{d}Q}{\mathrm{d}t} \biggr)_{1},
\label{eq:drho_rad}
\end{equation}
\begin{equation}
\biggl( \frac{\mathrm{d} \Delta \rho}{\mathrm{d}t} \biggr)_\mathrm{ad} = \rho_\mathrm{e} \frac{v_\mathrm{r}}{H_\mathrm{p}} \delta,
\label{eq:drho_ad}
\end{equation}
where $\delta=\nabla_\mathrm{e}-\nabla_\mathrm{ad}$ and $v_\mathrm{r}$ is the radial velocity of the flux tube apex.  It is found (see Figure 4 in \citet{fan_1996}) that in the lower one-third of the convection zone, d$\Delta \rho/$d$t$ is dominated by contributions from radiative heating.  This term is only dependent on the properties of the background plasma environment, and is therefore the same for every flux tube.  In the remaining two-thirds of the convection zone above $\approx$$0.80$\,R$_{\odot}$, the evolution of the flux tube can be described as essentially adiabatic, as the heating rate [$(\mathrm{d}Q/\mathrm{d}t)_{1}$] decreases with height and $\delta$ and $v_\mathrm{r}$ increase with height.  As will be shown later in Section \ref{sec:dynamics_rad}, the inclusion of the radiative heating term to the TFT energy equation has significant implications for the dynamic evolution and rise times of the flux tube.      

In Sections \ref{sec:dynamics_rad} and \ref{sec:emergence_properties} we will compare the results of the simulations calculated in Articles 1 and 2, where we assume that the flux tube evolves adiabatically, with the new simulations as described here in Section \ref{sec:model}.  For simplicity, we will often refer to the simulations of Articles 1 and 2 as case AB (adiabatic).  We will refer to the non-adiabatic simulations with the addition of radiative diffusion to the TFT energy equation as case RD.      

\section{Flux Tube Dynamics}
\label{sec:dynamics_rad}

\subsection{Flux Tube Morphology}
\label{sec:morph}

\begin{figure}
\begin{center}$
\begin{array}{ccc}
\includegraphics[scale=.20]{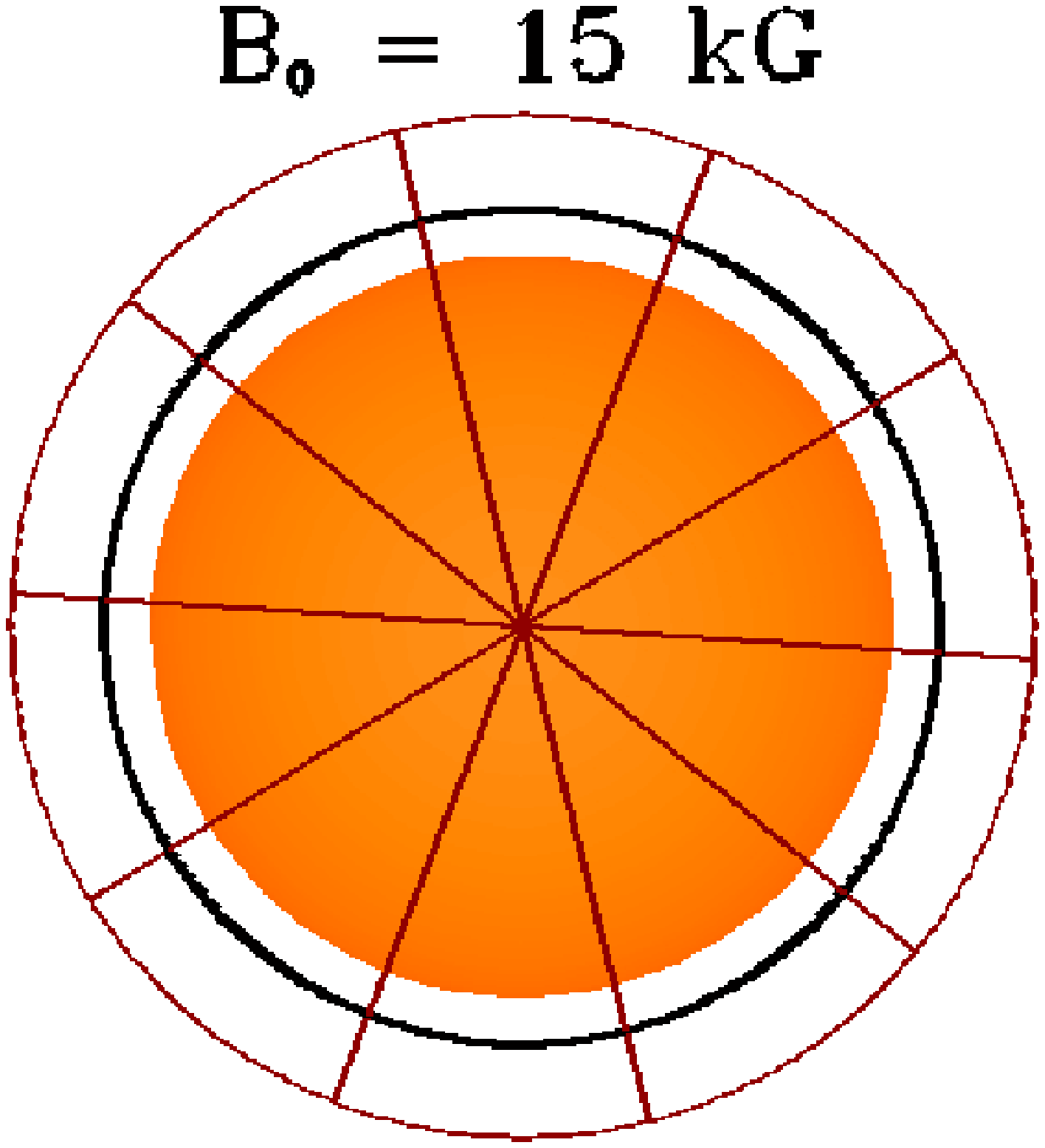} 
\includegraphics[scale=.20]{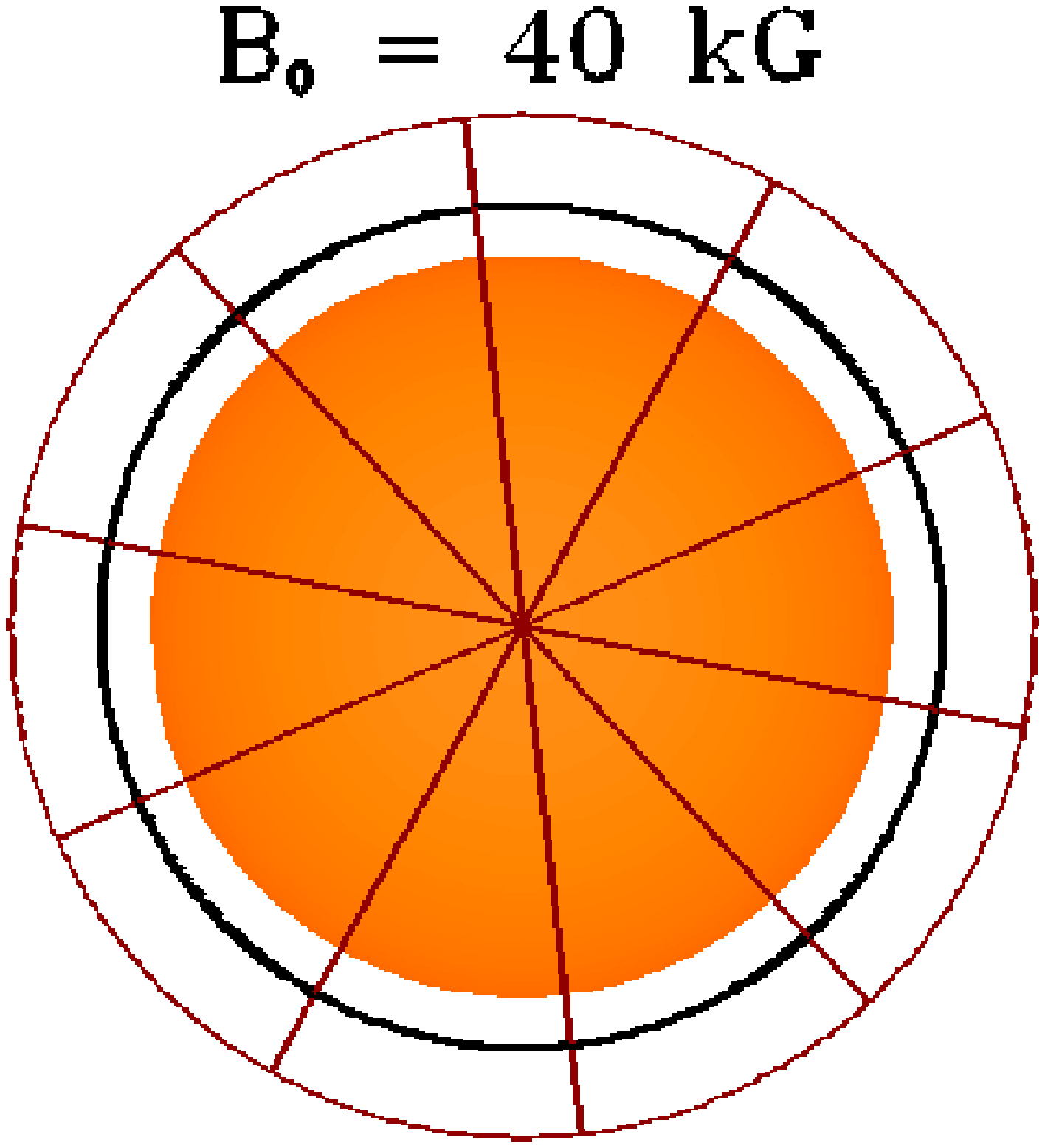} 
\includegraphics[scale=.20]{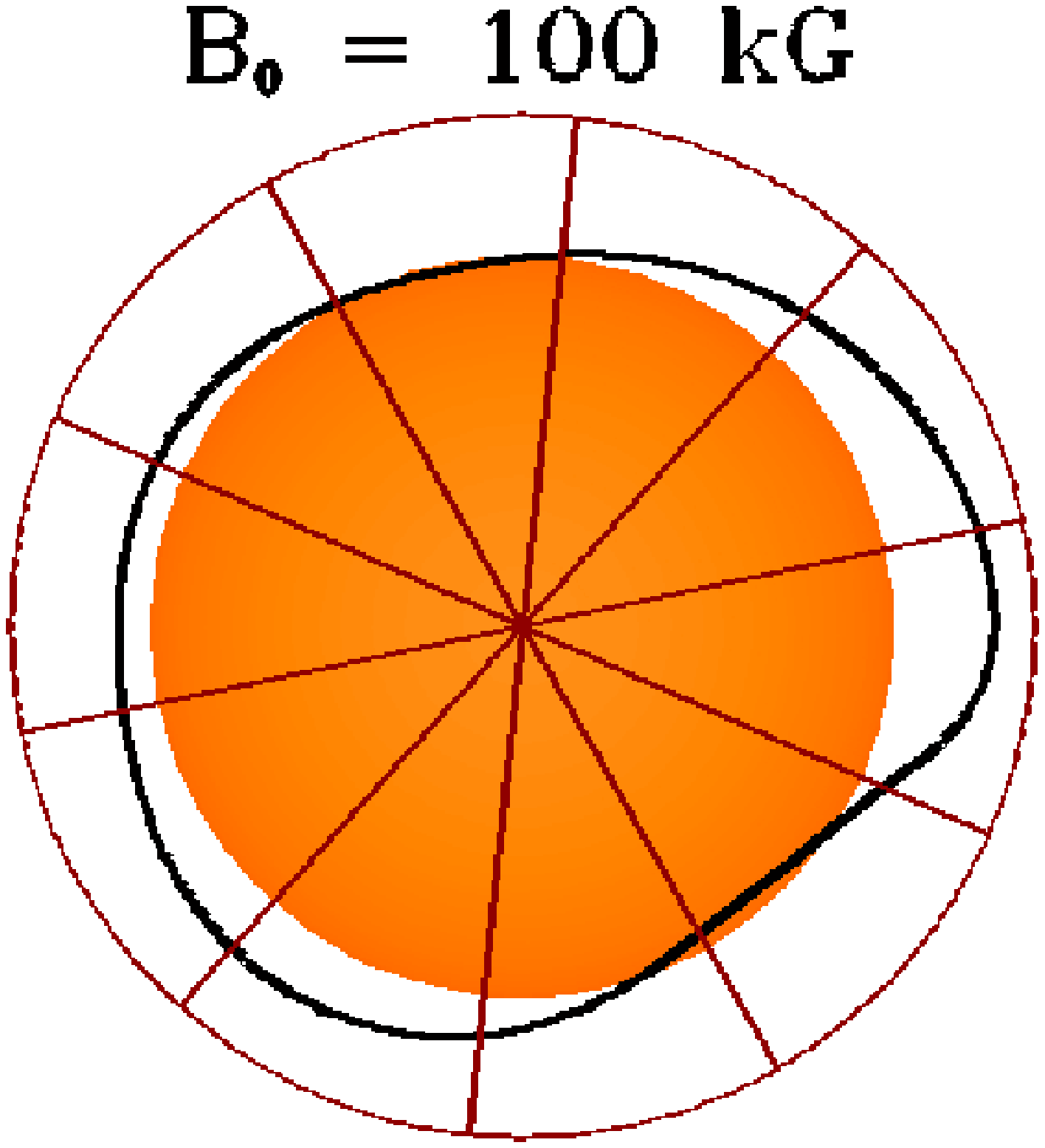} \\
\includegraphics[scale=.20]{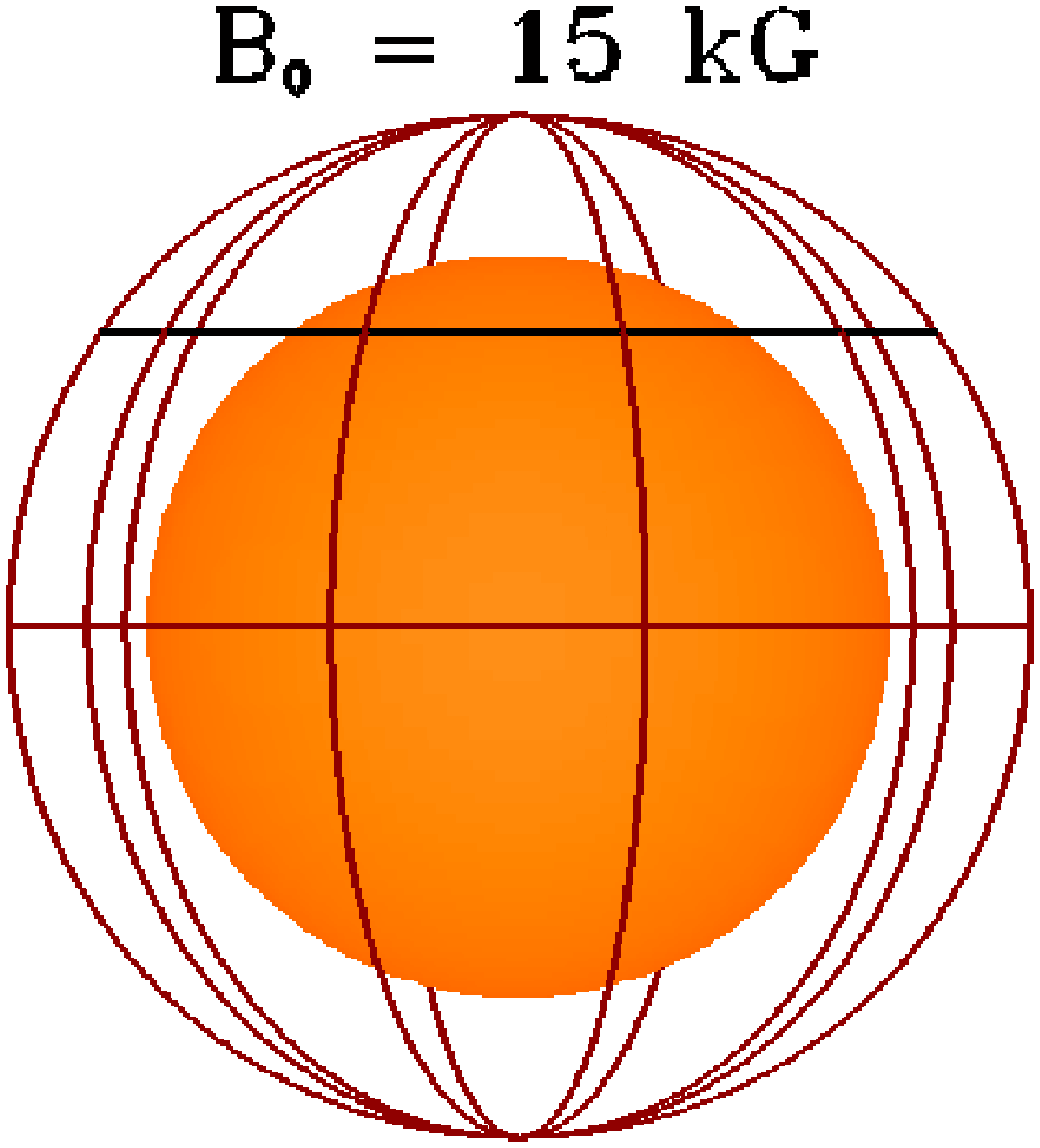} 
\includegraphics[scale=.20]{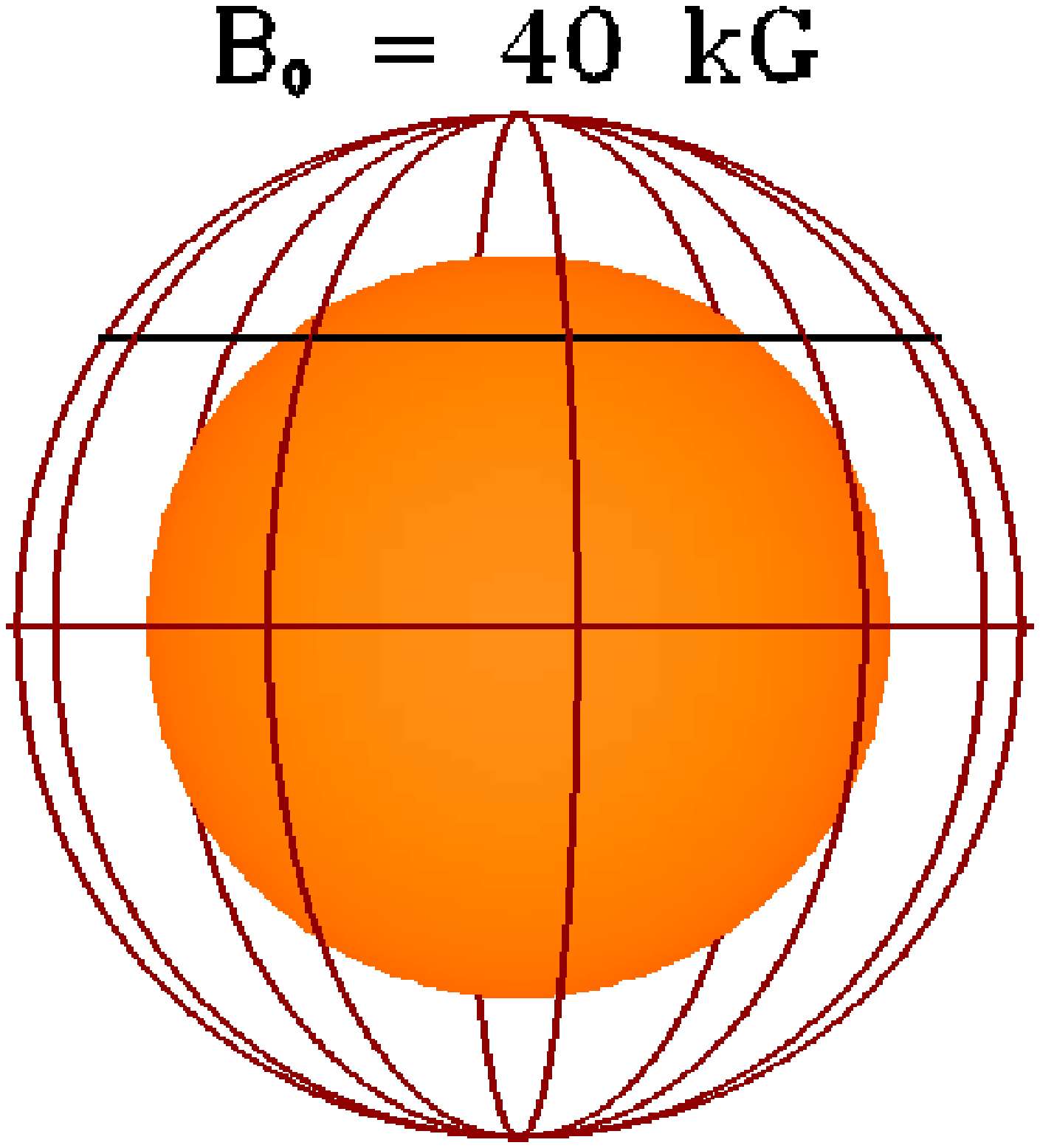} 
\includegraphics[scale=.20]{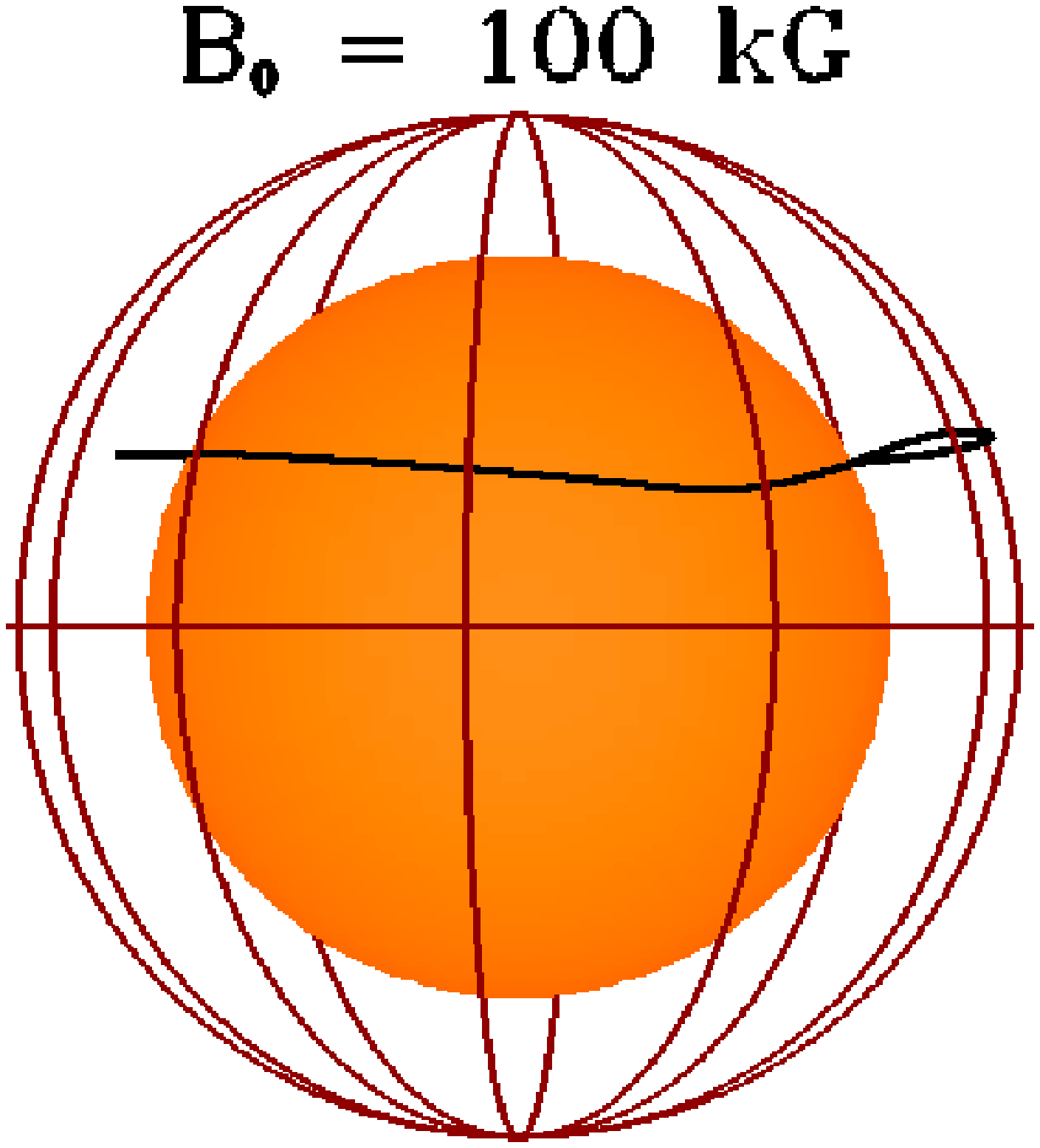}
\end{array}$
\end{center}
\vspace{-.06\textwidth}
\caption{Polar (top) and equatorial (bottom) view of flux tubes once some portion of the tube has reached the simulation upper boundary.  These $10^{22}$ Mx, $\theta_{0}=8^{\circ}$ flux tubes are allowed to evolve in the absence of convection, but with the addition of heating due to radiative diffusion.  The orange sphere has a radius of 4.9\e{10} cm.   The addition of radiative heating forces flux tubes to drift upward as a whole and away from the overshoot region.  Only 100 kG flux tubes of $\theta_{0} \le 8^{\circ}$ have footpoints that anchor in the overshoot region.}
\label{fig:snapshot_noconv_rad}
\end{figure}

When flux tubes from our simulations evolve adiabatically in the absence of convection, rising buoyant loops develop solely as a result of the non-linear growth of the magnetic buoyancy instability, as discussed in Article 1.  The troughs of these rising loops penetrate into the overshoot region where they remain anchored for the duration of the flux emergence process.  However, when heating due to radiative diffusion is considered, only 100 kG flux tubes of $\theta_{0} \le 8^{\circ}$ are capable of anchoring in the overshoot region (see Figure \ref{fig:snapshot_noconv_rad}).  

The term $(\mathrm{d}Q/\mathrm{d}t)_{1}$ acts to heat the flux tube uniformly (\textit{i.e.} uniformly for flux tube portions at the same distance [$r$]), especially near the base of the convection zone where the divergence of radiative heat flux from the external plasma environment is the greatest.  This uniform heating increases the flux tube's density deficit early in its rise, thereby increasing the buoyancy of the flux tube in the lower convection zone.  Only 100 kG flux tubes of $\theta \le 8^{\circ}$ develop undular magnetic buoyancy instabilities that grow fast enough to allow the troughs of at least one of their rising loops to penetrate into the overshoot region.  Magnetic buoyancy instabilities develop more slowly in weak magnetic-field-strength flux tubes.  Flux tubes of $\le$60 kG develop a uniform buoyancy quicker than the growth of the magnetic buoyancy instabilities, resulting in a nearly axisymmetric rise.  This is shown especially well for a 15 and 40 kG flux tube in Figure \ref{fig:snapshot_noconv_rad}.

\begin{figure}
\begin{center}$
\begin{array}{ccc}
\includegraphics[scale=.20]{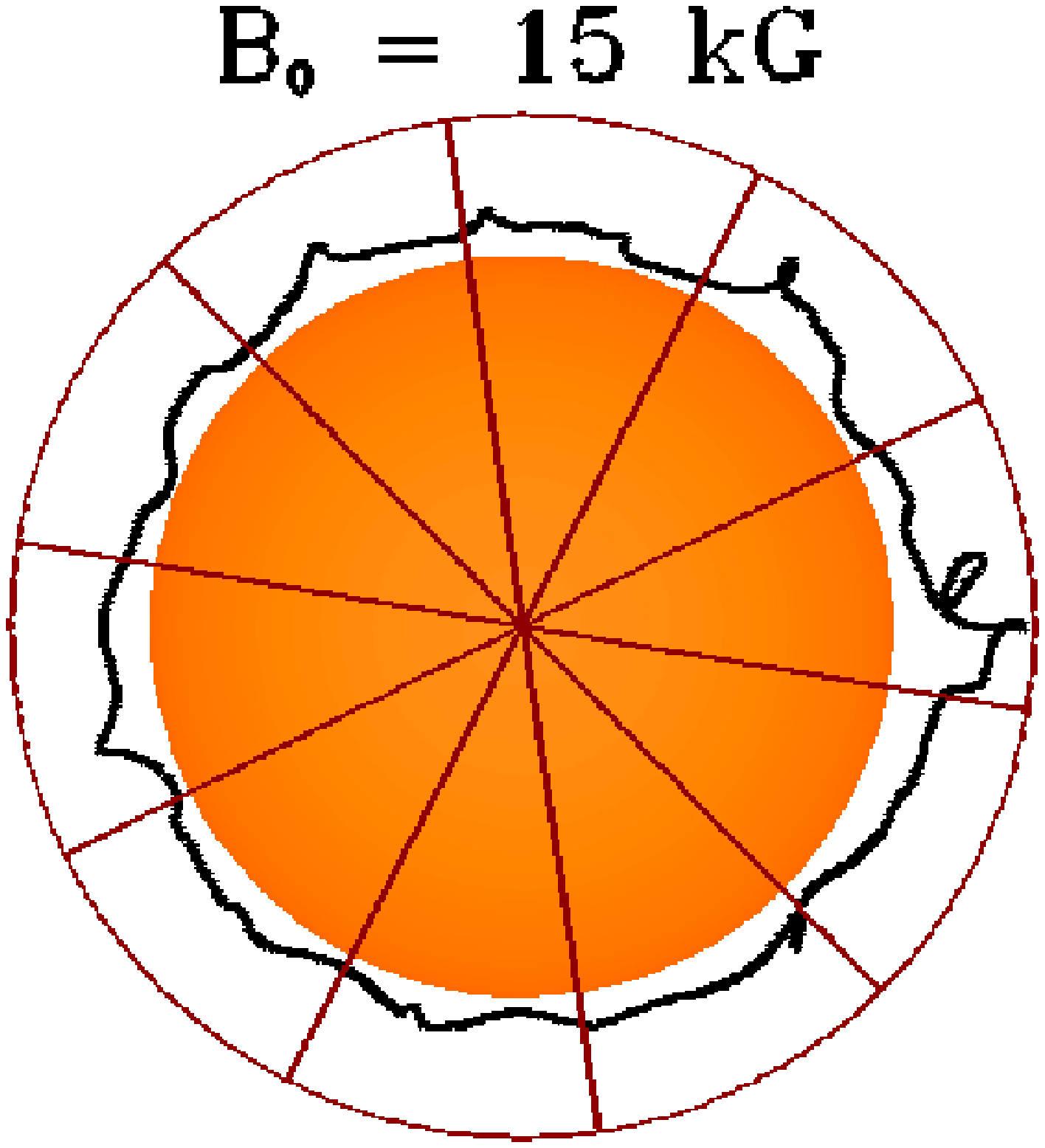} 
\includegraphics[scale=.20]{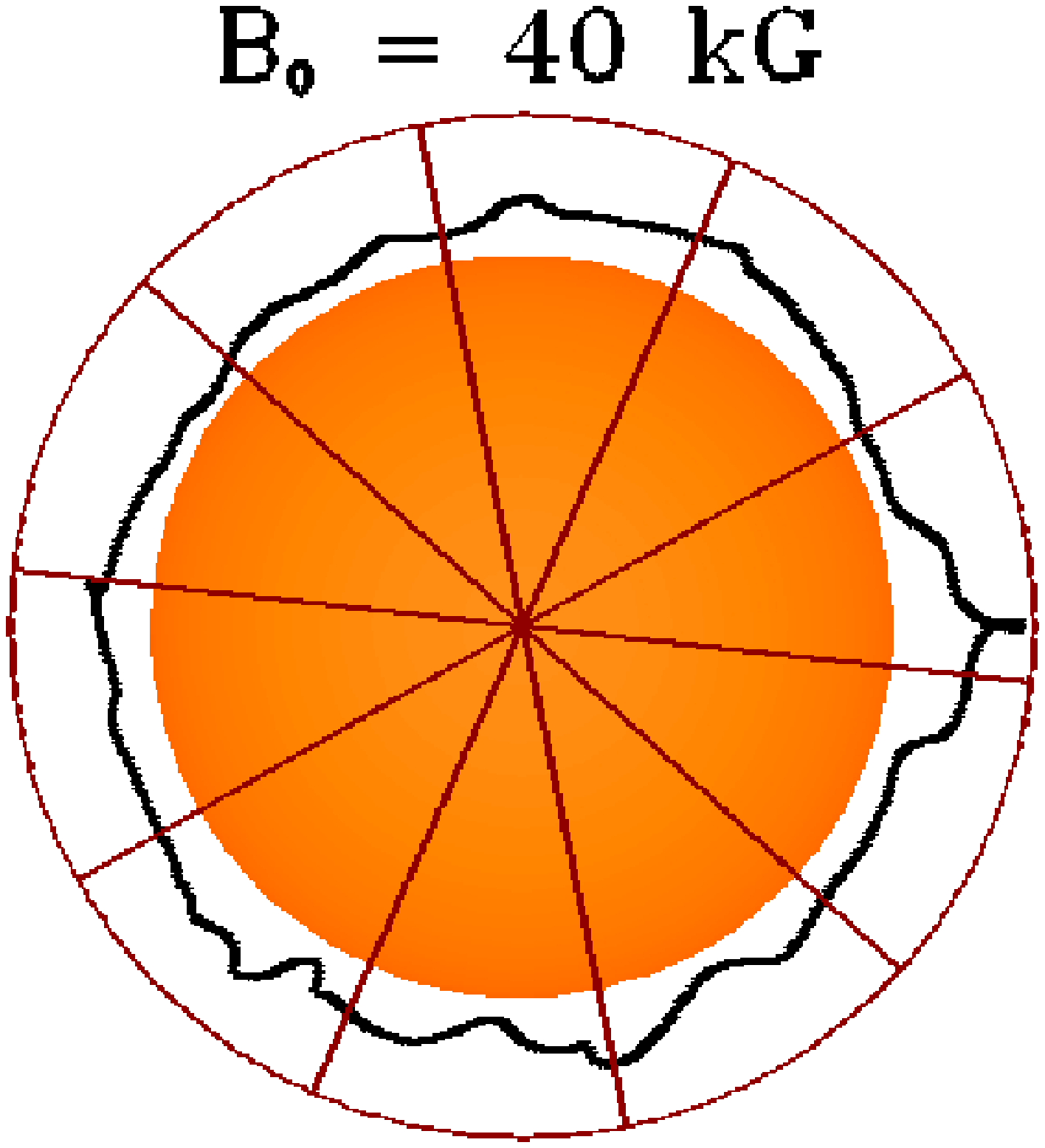} 
\includegraphics[scale=.20]{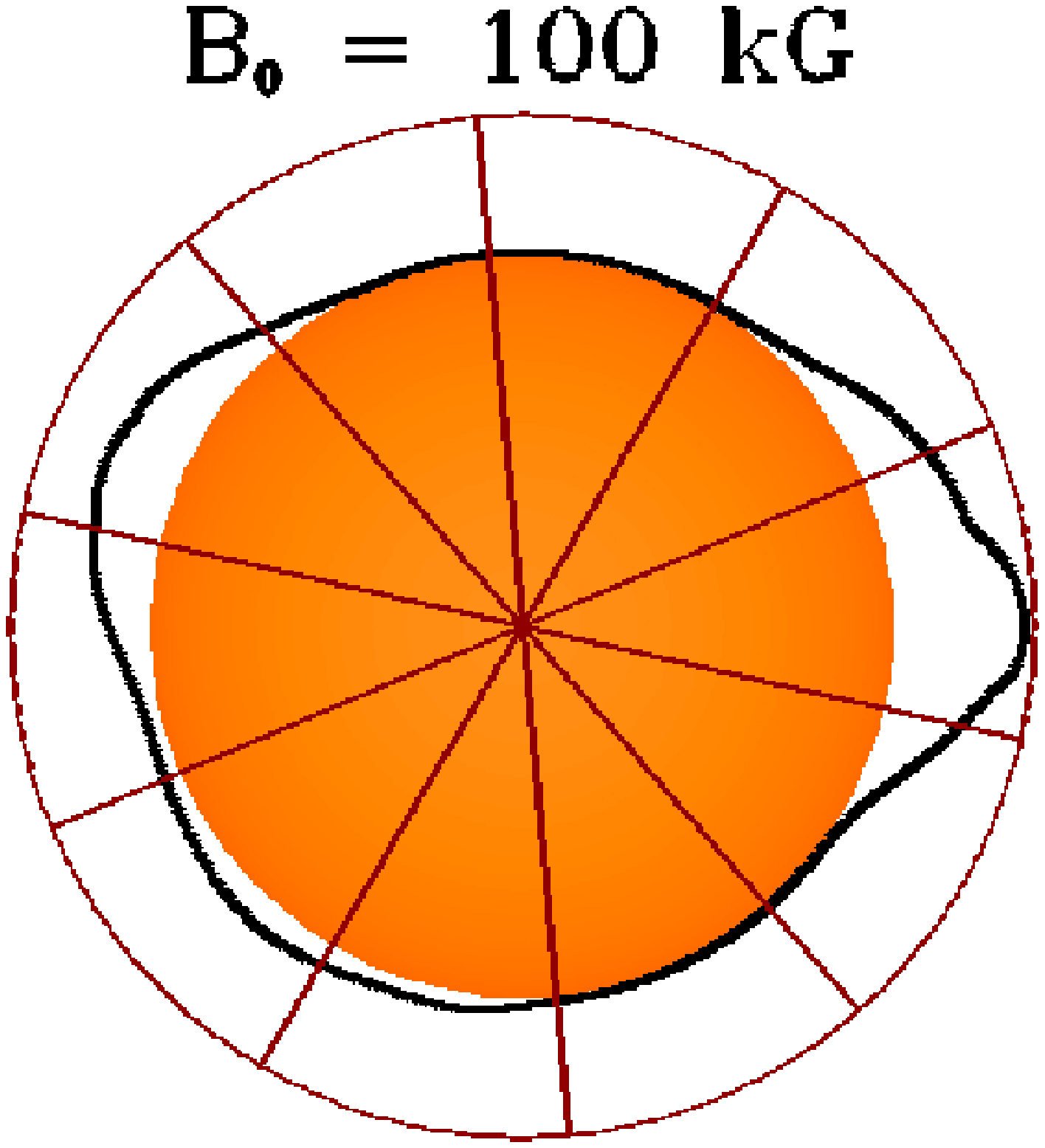} \\
\includegraphics[scale=.20]{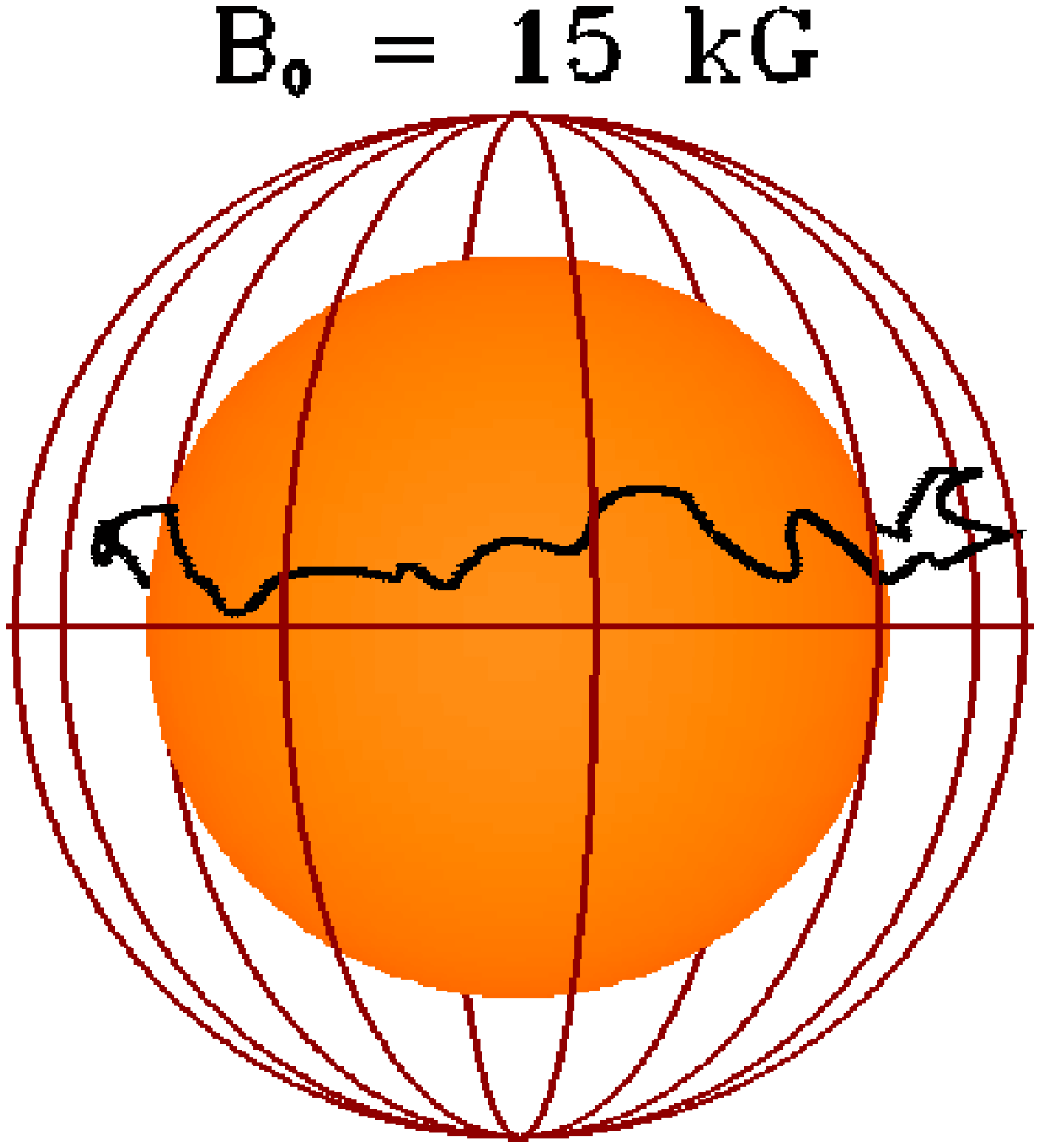} 
\includegraphics[scale=.20]{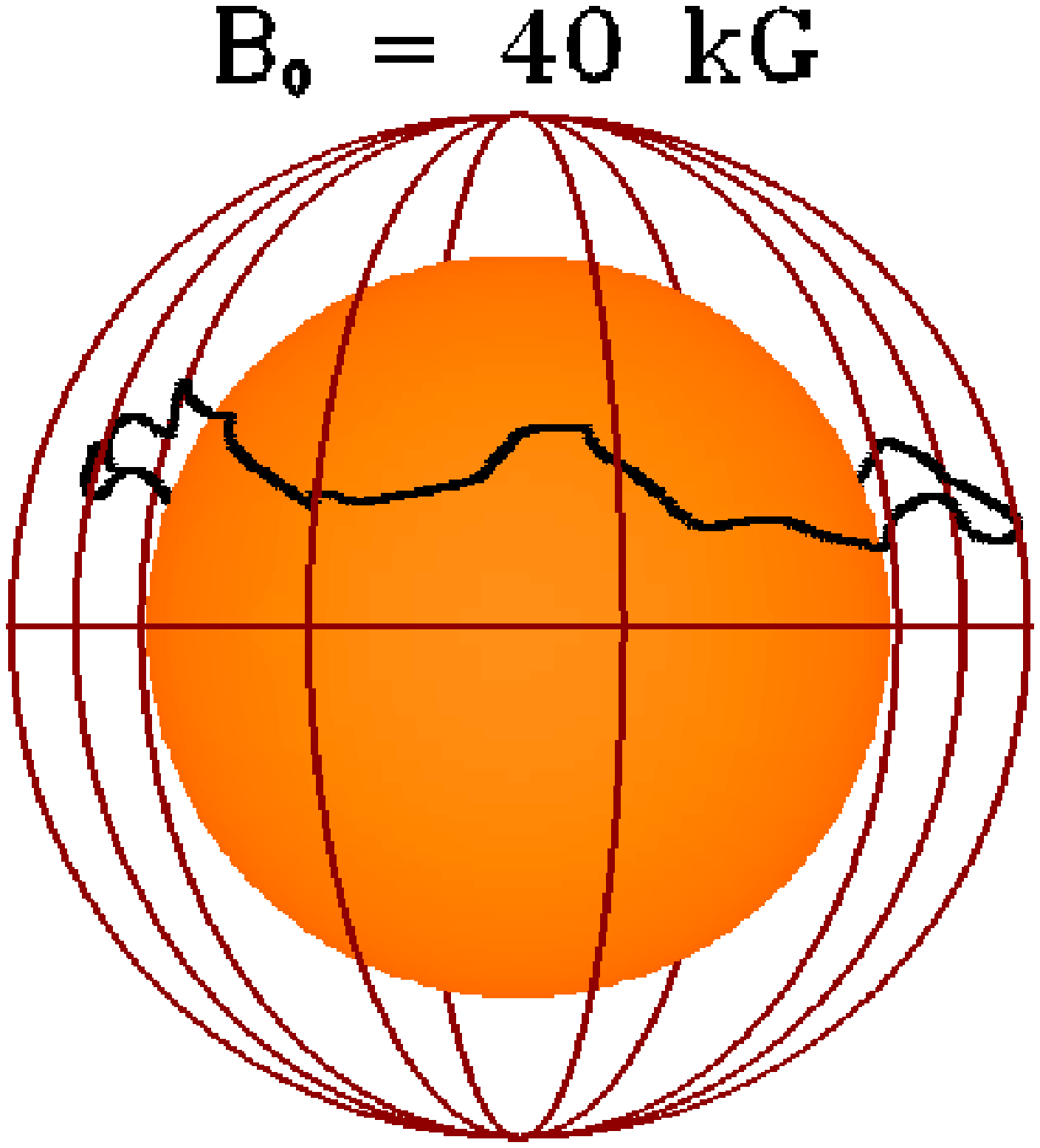} 
\includegraphics[scale=.20]{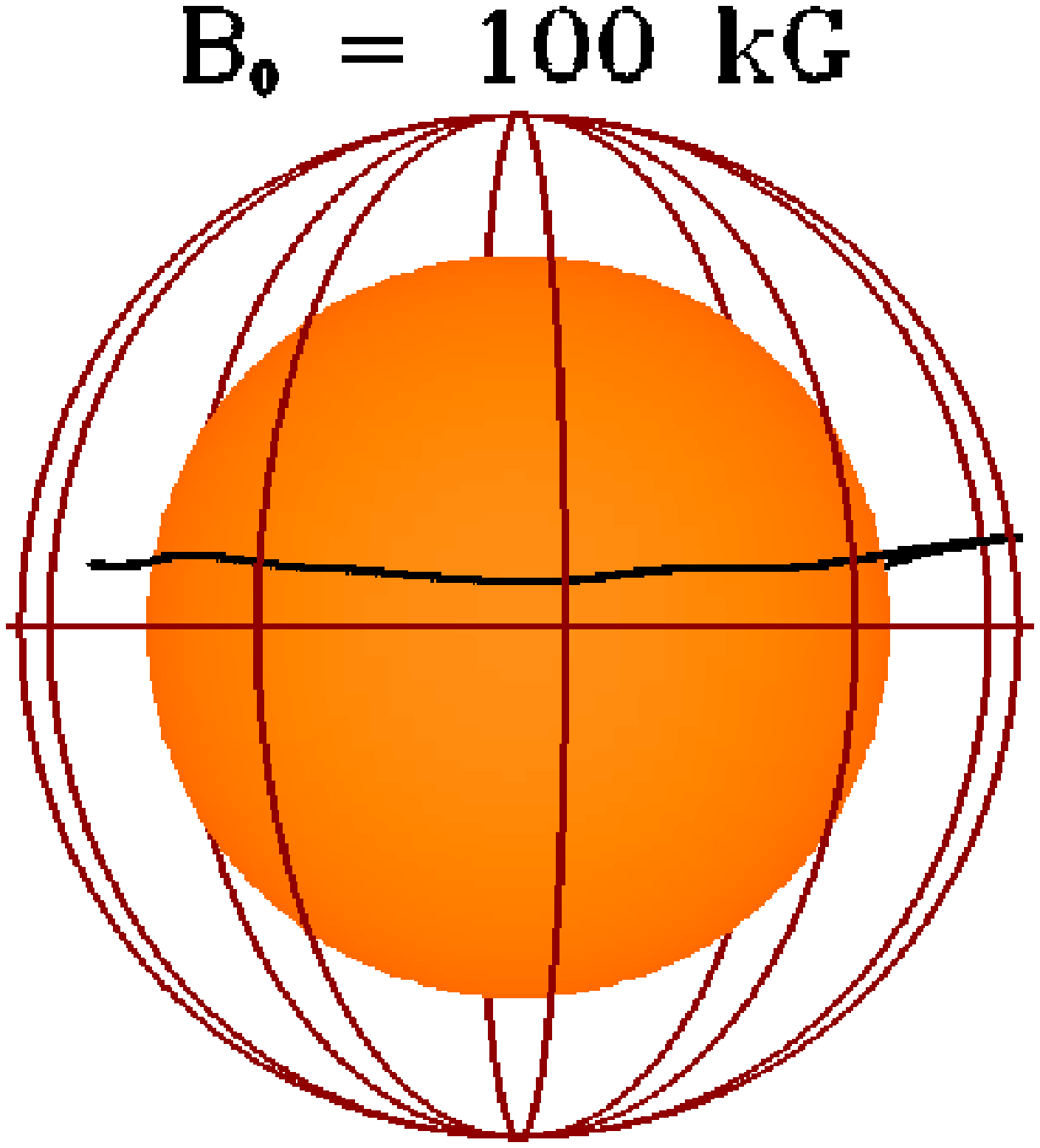}
\end{array}$
\end{center}
\vspace{-.06\textwidth}
\caption{Polar (top) and equatorial (bottom) view of flux tubes once some portion of the tube has reached the simulation upper boundary.  These $10^{22}$ Mx, $\theta_{0}=8^{\circ}$ flux tubes are the same as in Figure \ref{fig:snapshot_noconv_rad}, except the flux tube is subjected to the external convective flow.  In all cases, the image has been rotated such that the flux tube apex is on the right, and at the three o$'$clock position if looking down from the north solar Pole.  These flux tubes are referred to here as case RD flux tubes, indicating that radiative diffusion and convection are included in the simulations. Convection prevents the severe poleward slippage of the flux tube as depicted in Figure \ref{fig:snapshot_noconv_rad}, and weaker magnetic-field-strength flux tubes are more susceptible to deformation by convection.}
\label{fig:snapshot_conv_rad}
\end{figure}

\begin{figure}
\centerline{\includegraphics[width=.8\textwidth,angle=90]{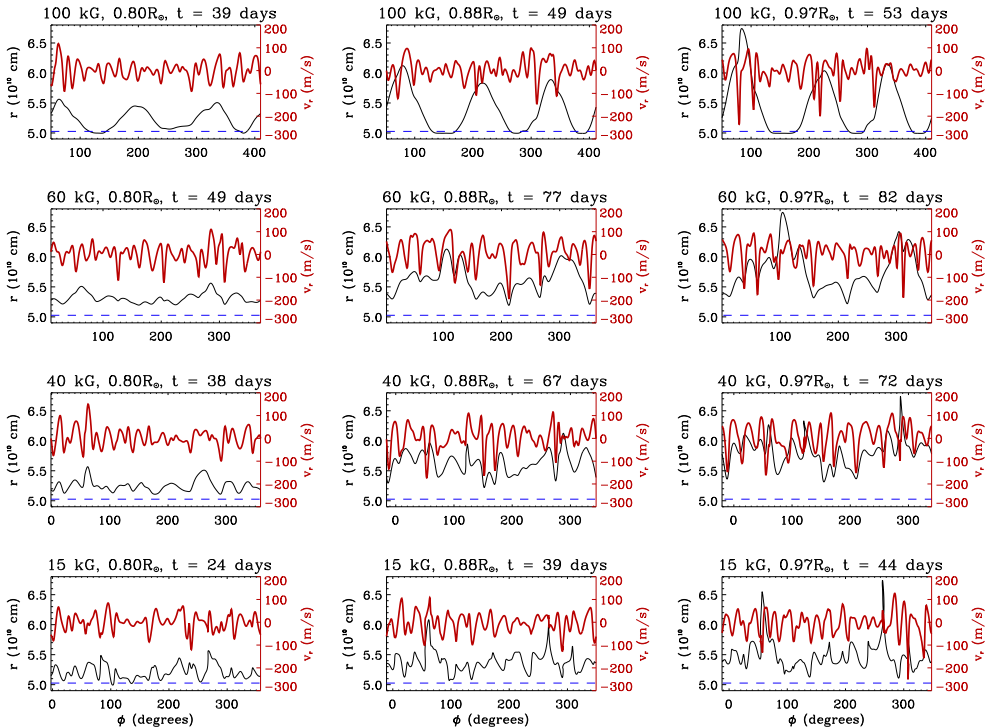}}
\caption{Flux tube radial distance from Sun center $r$ (black line), plotted with the external radial velocity experienced by the flux tube at the height $r$ of the flux tube segment (heavy red line), both as functions of the azimuthal angle [$\phi$].  Snapshots are for case RD flux tubes with $\Phi=10^{22}$ Mx, $\theta_{0}=15^{\circ}$, for $B_{0}$ decreasing from top to bottom.  All flux tubes shown are initialized at the same time, and therefore experience the same flow field initially.  The dashed line represents the base of the convection zone, below which is the stably stratified overshoot region.  These plots show the evolution of the flux tube once the apex has reached a height of (left) 0.80\,R$_{\odot}$, (middle) 0.88\,R$_{\odot}$, and (right) 0.97\,R$_{\odot}$, indicating that $\le$60 kG flux tubes subject to radiative heating do not anchor in the overshoot region.}
\label{fig:radial_vr_stuff_rad}
\end{figure}

When solar-like convective flows are included in the thin flux tube simulations (denoted as case RD flux tubes here because of the inclusion of radiative diffusion to the model), we find that 15 kG flux tubes are affected most strongly by convection, whereas 100 kG flux tubes only have moderate deformations resulting from the strongest convective downdrafts (see Figure \ref{fig:snapshot_conv_rad} and Figure \ref{fig:radial_vr_stuff_rad}).  This trend is the same as observed for case AB (\textit{i.e.} flux tube simulations with convection, but only adiabatic evolution) in Article 1.  The addition of convection also keeps flux tubes from suffering the severe poleward slippage shown in Figure \ref{fig:snapshot_noconv_rad}, especially at weaker magnetic-field strengths.  

Figure \ref{fig:radial_vr_stuff_rad} clearly shows that when convection is included, case RD flux tubes of $\le$60 kG do not anchor in the overshoot region.  Rather, due to heating from radiative diffusion, these flux tubes develop a density deficit very early in their evolution and begin to float away from the base of the convection zone before magnetic buoyancy instabilities of sufficient amplitude can set in to anchor the flux tube in the overshoot region.  They reach the middle of the convection zone where they are continually buffeted by convection until a buoyant loop reaches the simulation upper boundary.  All flux tubes of $\ge$80 kG do anchor when convection is included, and their footpoints continue to drift out of the overshoot region as the tube evolves due to the uniform heating supplied by $(\mathrm{d}Q/\mathrm{d}t)_{1}$.  This anchoring is facilitated by perturbations to the flux tube provided by the convective flow field, initiating magnetic buoyancy instabilities of large enough amplitude to anchor the flux tube footpoints before the tube as a whole can drift away from the lower convection zone.  These convective perturbations are larger in amplitude than the superposition of Fourier modes applied to the flux tubes to promote undular magnetic buoyancy instabilities when convection is not present.     

It is worthwhile to briefly mention the effects of mean flows from our ASH convection simulation on flux tube evolution.  As mentioned in Article 1, differential rotation present in the convection simulation tends to boost the azimuthal velocity of the flux tube mass elements in the prograde direction.  We do not expect this to have a significant effect on the flux tube properties discussed in this article.  The kinetic energy of the mean meridional flow present in our ASH simulation is approximately two orders of magnitude smaller than the kinetic energy of convection and differential rotation.  For the range of flux tubes that we consider in this study, it is unlikely that meridional circulation contributes much to flux tube evolution.

\subsection{Convection vs. Magnetic Buoyancy}
As in Article 1, we compare the magnitude of the drag force to the magnetic buoyancy force acting on the flux tube to understand their relative importance on flux tube evolution for case RD simulations.  Following Equations (6) and (7) from Article 1, for the drag force to dominate the buoyancy force in the radial direction:
\begin{equation}
v_\mathrm{cr} > v_\mathrm{a} \bigg( \frac{a}{H_\mathrm{p}} \bigg)^{1/2},
\label{eq:compare2}
\end{equation}
\textit{i.e.} the convective flow speed [$v_\mathrm{cr}$] needs to be greater than the Alfv\'en speed [$v_\mathrm{a}=B/(4\pi\rho_\mathrm{e})^{1/2}$] multiplied by the term $(a/H_\mathrm{p})^{1/2}$.    

\begin{figure}
\centerline{
\includegraphics[scale=0.43,trim=0.55cm 0cm 0.8cm 0cm,clip=true]{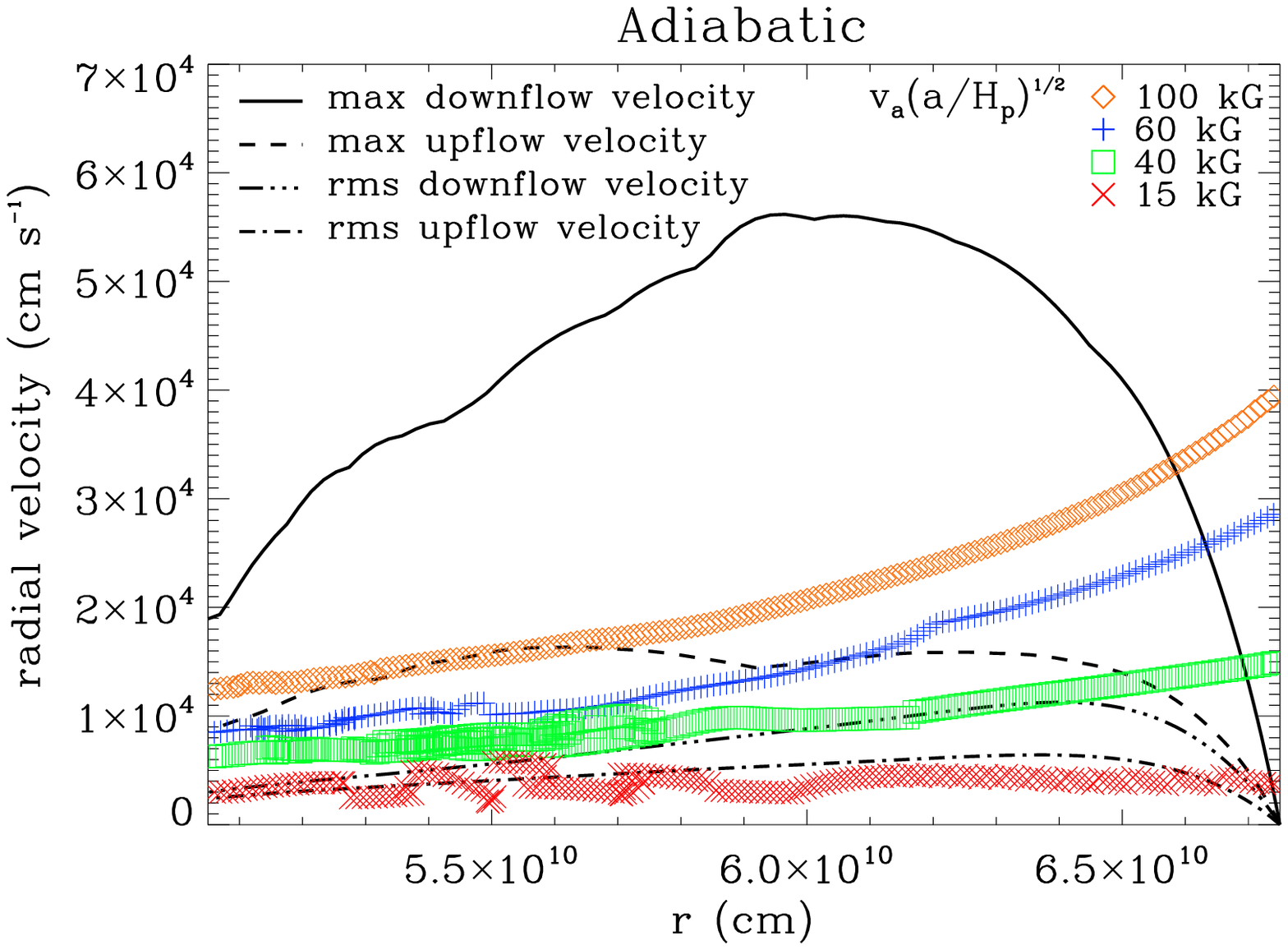} 
\includegraphics[scale=0.43,trim=1.1cm 0cm 0.8cm 0cm,clip=true]{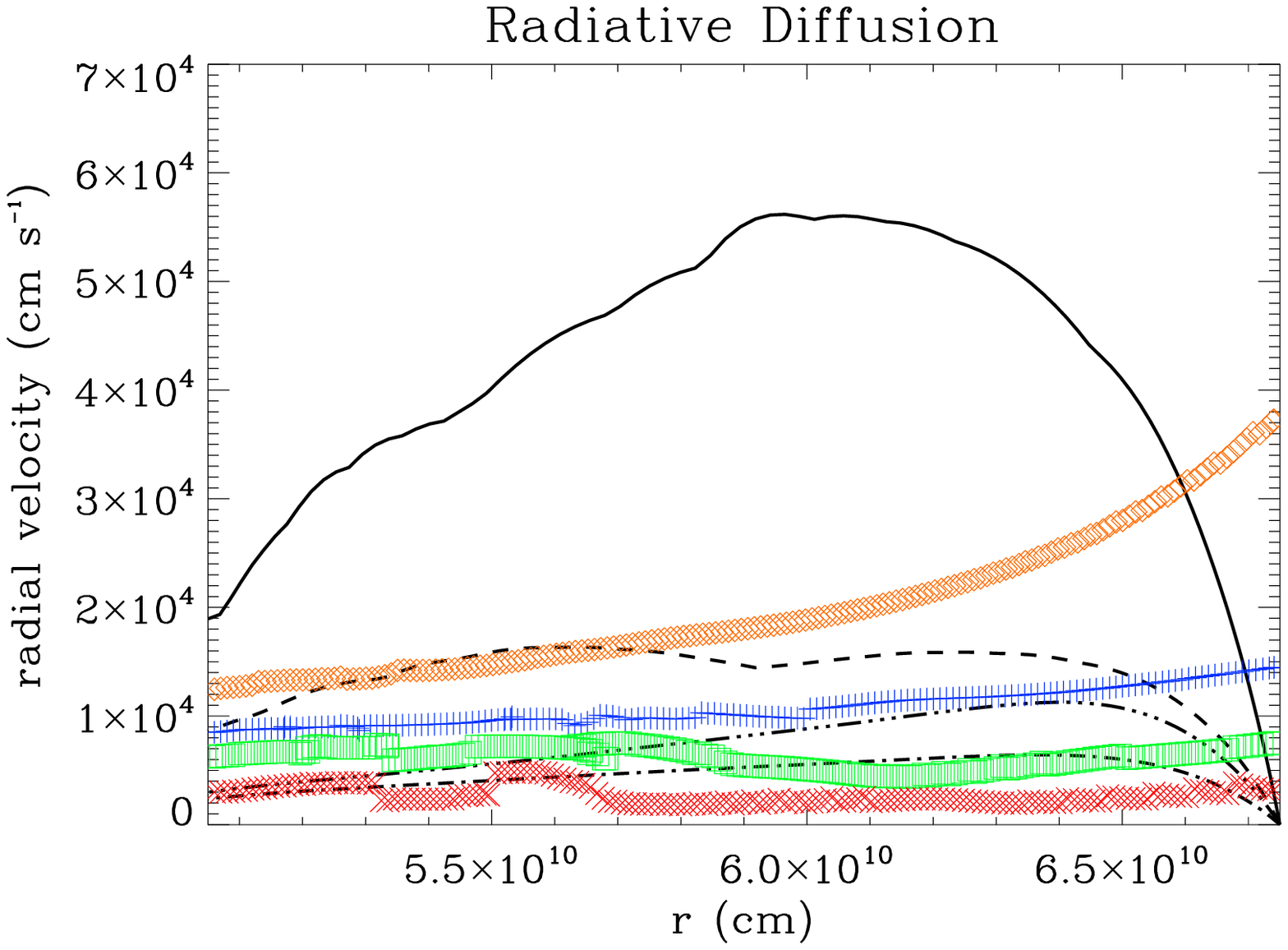}}
\caption[\quad The effects of convection vs. magnetic buoyancy on flux tubes subject to radiative heating]{The right-hand side of Equation (\ref{eq:compare2}) (colored symbols) plotted for flux tubes evolving in a convective velocity field assuming adiabatic evolution (case AB, left), and with the addition of radiative heating (case RD, right).  The right-hand side of Equation (\ref{eq:compare2}) decreases with decreasing magnetic field strength, therefore the top symbol curve represents 100 kG flux tubes (diamond), followed by 60 kG (plus sign) and 40 kG (square), with the bottom symbol curve representing 15 kG flux tubes (cross).  Also plotted are representative convective velocity field speeds in the radial direction  from the ASH simulation (black lines).  Flux tubes of $\le$60 kG are significantly more susceptible to convective influences when heating due to radiative diffusion is included.}
\label{fig:compare_alf_adia_rad}
\end{figure} 

In Figure \ref{fig:compare_alf_adia_rad}, we have plotted the right-hand side of Equation (\ref{eq:compare2}) for flux tubes of $\Phi=10^{22}$ Mx and $\theta_{0}=15^{\circ}$ for four different initial magnetic field strengths.  For comparison, the left panel shows the right-hand side of Equation (\ref{eq:compare2}) for case AB, and the right panel for case RD.  Each flux tube is subjected to the same flow field, and representative radial convective-flow speeds are also plotted.  Flux tubes of $B_{0}=100$ kG in both the left and right panels of Figure \ref{fig:compare_alf_adia_rad} are only affected by the strongest downflows, and the plots of $v_\mathrm{a}(a/H_\mathrm{p})^{1/2}$ at the apex of the loop are very similar.  However, at weaker initial magnetic field strengths, the quantity $v_\mathrm{a}(a/H_\mathrm{p})^{1/2}$ is decreased in the upper convection zone for case RD flux tubes as compared to case AB flux tubes.  Figure \ref{fig:compare_alf_adia_rad} shows that case RD flux tubes of $\le$60 kG are affected more strongly by convection than case AB flux tubes of the same initial magnetic field strength.  This occurs because the field strength of the flux tube apex becomes weaker for the case with radiative heating. 

\begin{figure}
\centerline{
\includegraphics[scale=0.384,trim=1.15cm 0cm 0.6cm 0cm,clip=true]{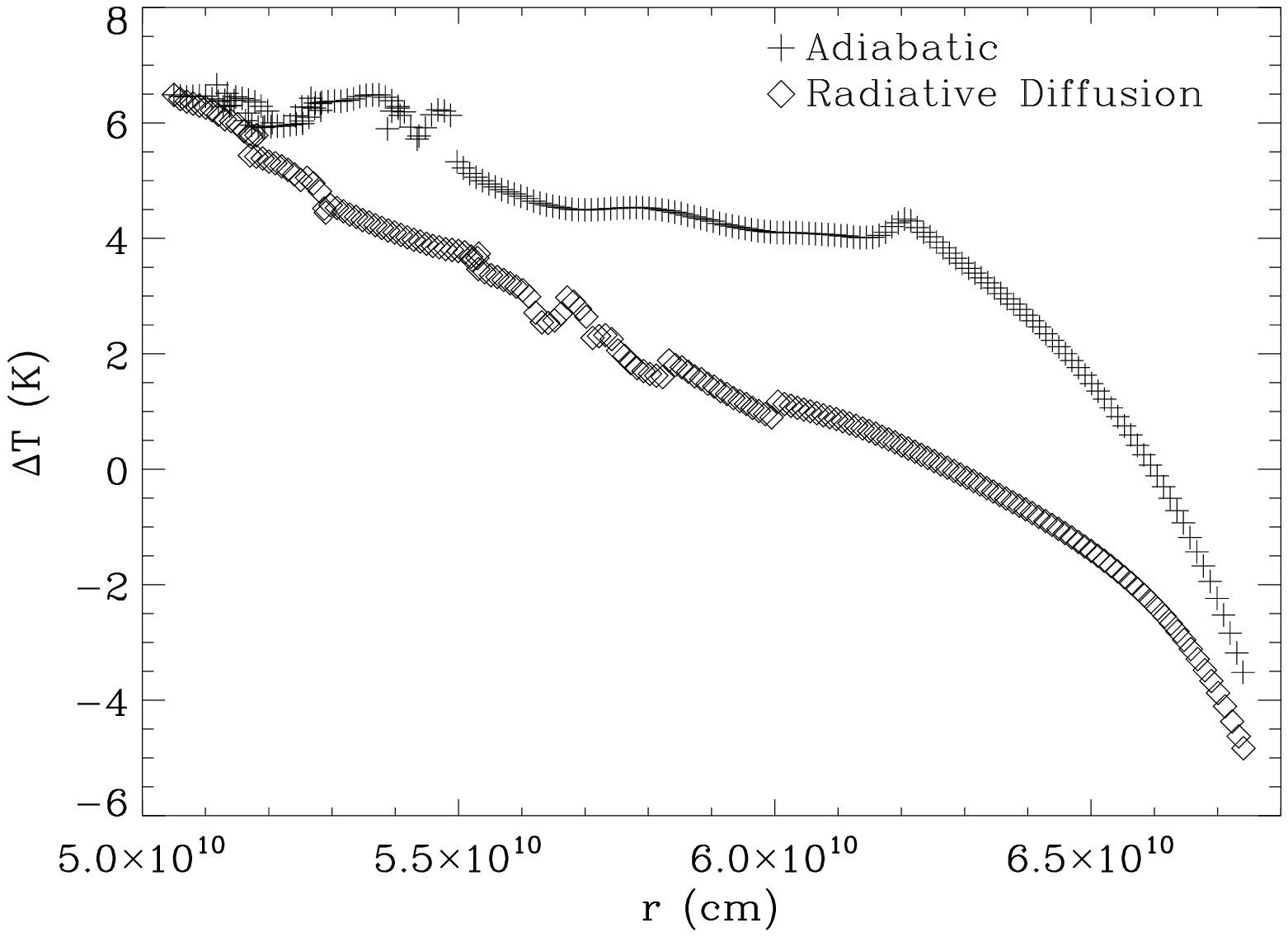}
\includegraphics[scale=0.384,trim=0.1cm 0cm 0.8cm 0cm,clip=true]{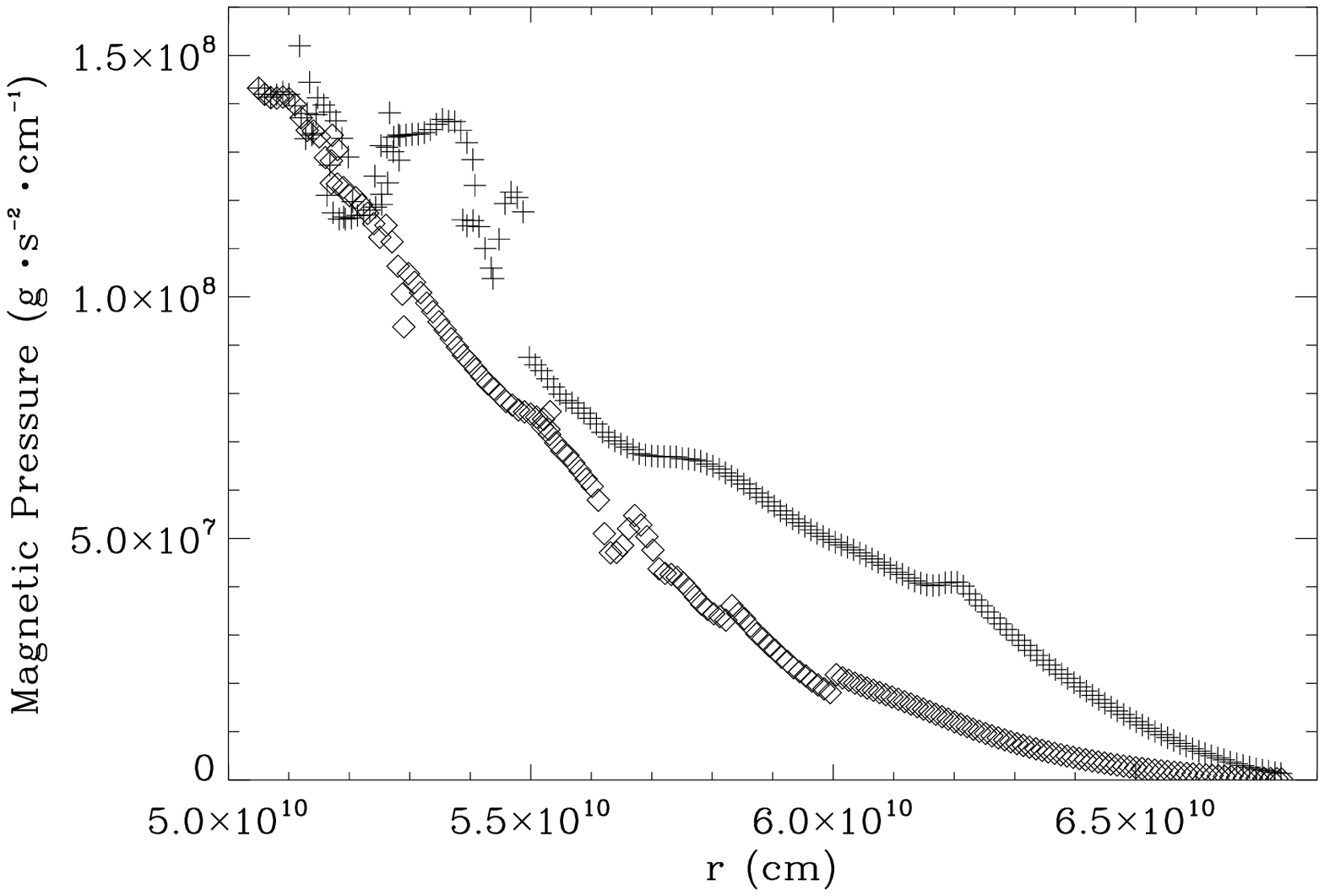}}
\caption{(Left) $\Delta T=T_\mathrm{e}-T$ and (Right) magnetic pressure at the apex of a flux tube as a function of height, where $\Phi=10^{22}$ Mx, $B_{0}=60$ kG, and $\theta_{0}=15^{\circ}$.  These quantities are shown for a case AB flux tube (plus symbols), and a case RD flux tube (diamond symbols), both rising through a turbulent solar-like convection zone.  Heating supplied by radiative diffusion increases the internal temperature of case RD flux tubes as compared to case AB flux tubes, resulting in a comparative increase in internal gas pressure which contributes to a decrease of magnetic pressure.}
\label{fig:apex_stuff}
\end{figure} 

Due to radiative diffusion in the lower convection zone, flux tubes experience an increase in their internal temperature.  The quantity $\Delta T=T_\mathrm{e}-T$ at the apex of a case AB and case RD flux tube, both subject to the same convective flows, is shown in the left-hand side of Figure \ref{fig:apex_stuff}.  In the lower portion of the convection zone, heating due to radiative diffusion increases the internal temperature of case RD flux tubes as compared to case AB flux tubes.  This results in a comparative increase in the internal gas pressure of case RD flux tubes, contributing to a decrease in the magnetic field strength so that the condition of pressure balance is fulfilled (Equation (\ref{eq:eqn_pbalance})).  The magnetic pressure at the apex of a case AB and case RD flux tube is also shown in the right-hand side of Figure \ref{fig:apex_stuff}.  Indeed, the magnetic pressure, and therefore the magnetic field strength at the apex of the case RD flux tube is less than it is for the case AB flux tube throughout the bulk of the convection zone.  This reduced magnetic field at the flux tube apex implies that the flux tube will be advected more strongly by convection.

\subsection{Rise Times}
\label{sec:rise_times}

Adiabatically evolving flux tubes spend the majority of their rise times at the base of the convection zone as the magnetic buoyancy instability grows.  In our simulations, convective flows provide perturbations to the flux tube that initiate the growth of these instabilities.  However, when radiative heating is included, the flux tube receives a kick start in its rise toward the surface, as the flux tube is heated uniformly by the non-zero divergence of radiative heat flux in the lower convection zone, thereby increasing its density deficit [$\Delta\rho=\rho_\mathrm{e}-\rho$] earlier on, speeding up its rise through the solar convection zone. 

\begin{figure}
\centerline{\includegraphics[scale=.55]{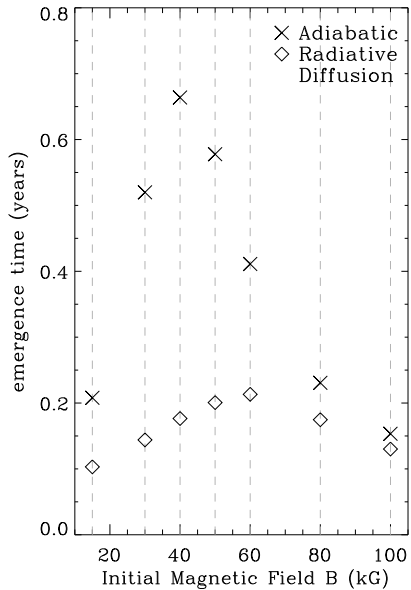}}
\caption{Average rise times for $\Phi=10^{22}$ Mx case AB (crosses), and case RD (diamonds) flux tubes.  Radiative heating reduces the average rise time of the flux tube in most cases.}
\label{fig:risetimes_rad_adia}
\end{figure}

The inclusion of radiative diffusion in the TFT energy equation reduces the flux tube rise times to $\le$0.2 years for all magnetic field strengths, as compared to $\le$0.7 years for flux tubes evolving adiabatically (see Figure \ref{fig:risetimes_rad_adia}).  The reduction is minimal for 80 and 100 kG flux tubes.  In this large-magnetic-field-strength regime, the average rise times are nearly the same because of the strong magnetic buoyancy and tension of the flux tubes, and because their undular magnetic buoyancy instabilities of order $m=1$ to $m=3$ grow quickly, facilitating anchoring in the overshoot region.   

There is a large reduction of rise times for case RD flux tubes in the $B_{0}=15-60$ kG regime as compared to case AB flux tubes: a difference of $\approx$$0.05-0.5$ years.  This reduction is a result of radiative heating contributing a significant increase to the buoyancy of the flux tube early in its rise.  Radiative diffusion also increases the internal pressure of the tube early on, forcing a reduction in its magnetic field strength, therefore it is more susceptible to convection.  

Case AB flux tubes also show a large spread in their average rise times, from $\approx$$0.15-0.7$ years.  This is a result of the complex interaction between the buoyancy and drag forces on the flux tube for various magnetic field strengths (see Articles 1 and 2).  Case RD flux tubes only exhibit a spread in rise times of $\approx$$0.1-0.2$ years.  The average rise times are brought closer together for all magnetic-field-strength flux tubes because they all experience the same heating rate at the same height in the deep convection zone. 

\section{Emergence Properties}
\label{sec:emergence_properties}

\subsection{Latitude of Emergence}
\label{sec:latem_rad}

\begin{figure}
\begin{center}$
\begin{array}{cc}
\includegraphics[scale=.75]{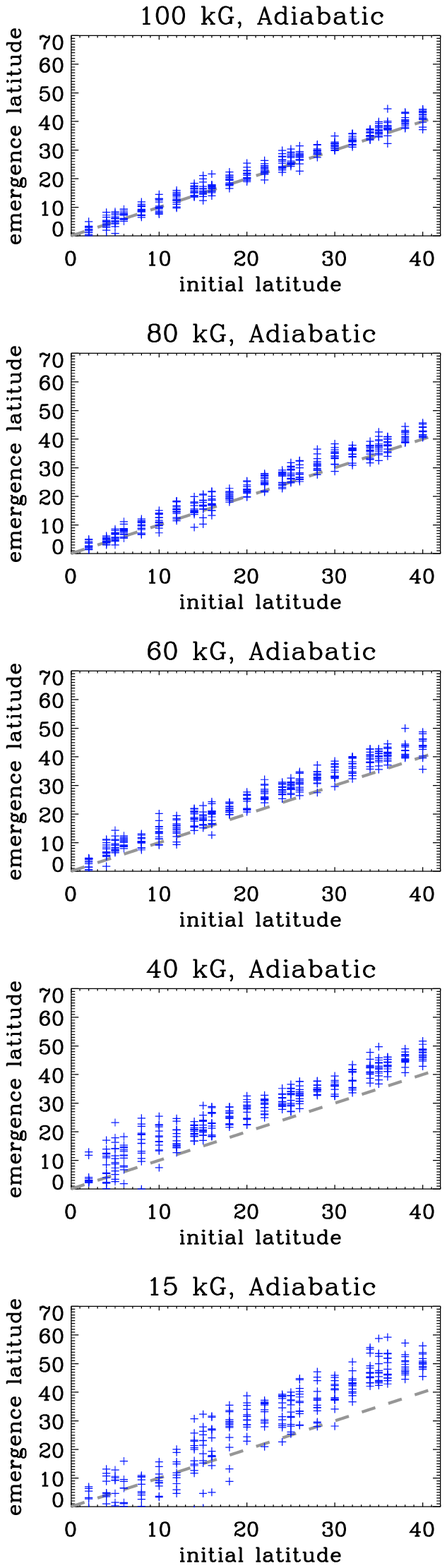} &
\includegraphics[scale=.75]{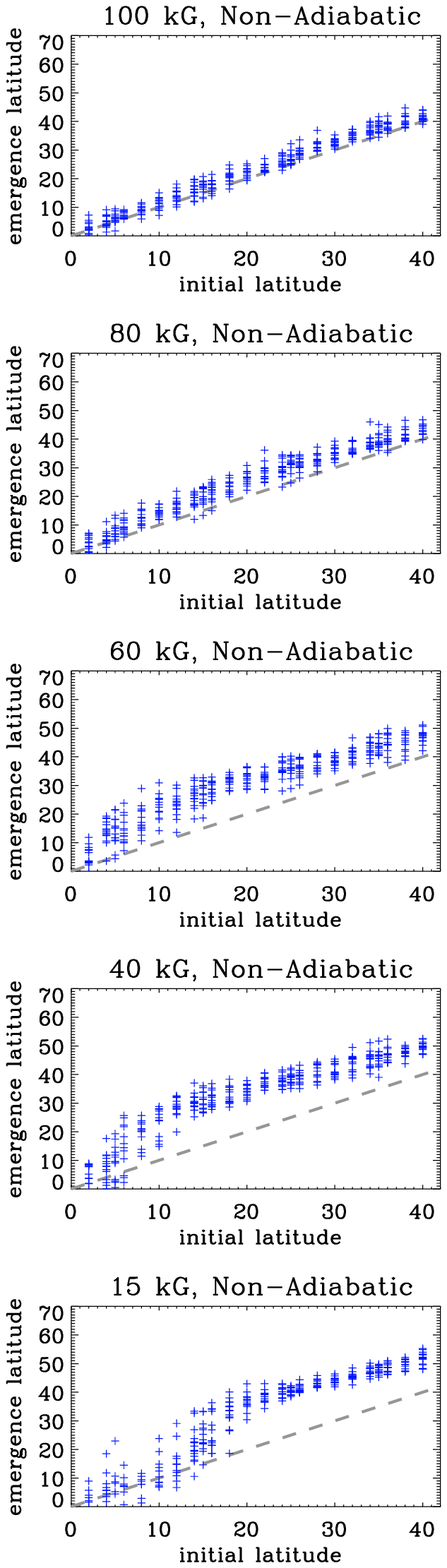}
\end{array}$
\end{center}
\caption{Initial latitude \textit{versus} emergence latitude of the apex of $10^{22}$ Mx flux tubes.  The left column shows the latitudinal deflection for case AB flux tubes, and the right column for case RD flux tubes (labeled as \emph{non-adiabatic} here).  Both axes are in units of degrees.  A dashed line indicates where the emergence latitude equals the initial latitude.  With the addition of radiative heating, middle-to-high latitude flux tubes of moderate to weak magnetic field strength deflect poleward.}
\label{fig:theta_rad}
\end{figure}

In the absence of convection, a flux tube will emerge radially with $\theta_\mathrm{em}\approx\theta_\mathrm{0}$ (emergence latitude $\approx$ initial latitude) if the outward component (away from the rotation axis) of the buoyancy force dominates the inward component (toward the rotation axis) of the Coriolis force \citep[\textit{e.g.}][]{choud_1987}.  Such a scenario occurs for flux tubes of large magnetic field strength, \textit{i.e.} 100 kG (see Articles 1 and 2).  Conversely, a flux tube rises more parallel to the rotation axis as the outward component of the buoyancy force is balanced by the inward component of the Coriolis force.  This happens as the initial magnetic field strength of the flux tube is reduced, \textit{i.e.} $B_{0}\le60$ kG, leading to a poleward deflection of the flux tube away from its initial latitude [$\theta_{0}$] that increases as $B_{0}$ decreases (see also Articles 1 and 2).

When convection is introduced, case AB (\textit{i.e.} adiabatically evolving) flux tubes of large $B_{0}$, and therefore larger buoyancy, still rise mostly radially (left panel of Figure \ref{fig:theta_rad}).  However, especially at weak field strengths of $\le$30 kG, the buoyancy force is reduced compared to the Coriolis force acting on the flux tube, forcing the apex of the flux tube to deflect poleward (\textit{i.e.} more parallel to the rotation axis).  This effect is responsible for the moderate latitudinal deflection of 15 kG flux tubes, especially at high latitudes (see Figure \ref{fig:theta_rad}, lower left panel).  As shown in Articles 1 and 2, the addition of solar-like convective flows prevents weak magnetic-field-strength flux tubes from deflecting poleward as severely as when convection is not included.  Convection also introduces a scatter in the emergence latitude values, which increases as $B_{0}$ decreases.  This happens because flux tubes become more susceptible to advection by convective flows as both the magnetic buoyancy and tension forces are reduced.   

Case RD (\textit{i.e.} radiative diffusion included) flux tubes of $80-100$ kG emerge with latitudes similar to case AB flux tubes of the same field strength (right panel of Figure \ref{fig:theta_rad}).  This is not surprising, as flux tubes in this field-strength regime do anchor in the overshoot region.  However, as the magnetic field strength of these flux tubes decreases, we begin to notice that case RD flux tubes tend to emerge at latitudes larger than case AB flux tubes, especially in mid-latitudes.  Due to the conservation of angular momentum, as case RD flux tubes rise as a whole away from the convection zone base, a retrograde flow of plasma inside the flux tube is enhanced, increasing the Coriolis force acting on the now rising flux ring, deflecting the case RD flux tubes more poleward.  At low latitudes for case RD flux tubes of $\le$60 kG, convective flows are able to keep the flux tube from slipping poleward, and there are no latitudinal zones void of flux emergence, as shown for case AB flux tubes in Article 1.

\subsection{Tilt Angles}
\label{sec:tilt_rad}
Solar active regions tend to emerge with their leading polarity (in the direction of solar rotation) closer to the Equator than the following, such that a line drawn between the center of the two bipolar regions will be tilted with respect to the East--West direction \citep{hale_1919}.  Here, the tilt angle is computed as the angle between the tangent vector at the apex of the emerging flux tube (once it has reached the top of the simulation domain), and the local East--West direction.  We define a positive sign of tilt as a clockwise/counter-clockwise rotation of the tangent vector away from the East--West direction in the Northern/Southern hemisphere, consistent with the direction of the observed mean tilt of active regions.  If the magnitude of the tilt angle exceeds 90$^{\circ}$, then the active region violates Hale's Law \citep{hale_1919}, possessing the \emph{wrong} leading polarity in the direction of solar rotation for that particular hemisphere.

To align our statistical results with those obtained from observations, unless otherwise stated, tilt angles derived from our flux tube simulations that do not fall in the range [-$90^{\circ}$, $90^{\circ}$] are shifted to be brought back into this interval, thereby losing information about anti-Hale tilt angles.  This approach is usually what is done for tilt-angle statistical studies of active regions on the Sun, starting with \citet{hale_1919} \citep[\textit{e.g.}][]{wang_sheeley_1989,fisher_1995,howard_1996,dasi_2010,stenflo_2012,li_2012,mcclintock_2013}.  If the leading polarity of the bipolar region in the direction of solar rotation, regardless of its sign, is closer to/farther from the Equator than the following, it will then be identified as a positive/negative tilt in the range [$0^{\circ}$, $90^{\circ}$]/[-$90^{\circ}$, 0$^{\circ}$].  This approach is different from what was done in Articles 1 and 2, where tilt angles were in the range of [-180$^{\circ}$, 180$^{\circ}$].  In retrospect, we feel that shifting the tilt angles to fall in the [-90$^{\circ}$, 90$^{\circ}$] range is probably a better diagnostic for comparison with observations. 

We perform numerous diagnostics on the tilt angles of our simulated flux tubes in an attempt to better constrain the magnetic field strength at which the solar dynamo may be operating, as was the topic of Article 2.  Also, we wish to identify the effect radiative diffusion has on tilt angle trends, comparing the results of the $10^{22}$ Mx case RD flux tube simulations calculated for this article with the case AB flux tubes from Articles 1 and 2.  

\subsubsection{The Joy's Law Trend}
\label{sec:joy_law}
Known as Joy's Law, the average tilting behavior of emerging-flux regions tends to increase in magnitude as the latitude of emergence increases \citep[\textit{e.g.}][]{hale_1919}.  Many authors have recovered Joy's Law from observations, however there is generally no agreed upon common method to obtain an empirical Joy's Law equation \citep[\textit{e.g.}][]{fisher_1995,dasi_2010,stenflo_2012,li_2012,mcclintock_2013}.  Within this section, we will employ two such methods in order to compare our simulation results to those of the Joy's Law trend recovered from observations.  

For our first method (Method 1) of obtaining an empirical equation for the Joy's Law trend from our simulations, we assume that the tilt angle increases linearly with increasing emergence latitude.  We perform a linear least-squares fit of the tilt angle as a function of emergence latitude following $\alpha = m_\mathrm{A} \theta$, where $\alpha$ and $\theta$ represent the tilt angle and emergence latitude respectively, both in units of degrees, of our simulated flux tubes once the apex has reached the simulation upper boundary, and $m_\mathrm{A}$ is the slope (unitless) of the best-fit line.  The slopes $m_\mathrm{A}$ of these best-fit lines along with their uncertainties are reported in columns 2 and 3 of Table \ref{tbl:tilts_radheat}.  Note that the values of $m_\mathrm{A}$ for case AB flux tubes are different from those reported in Article 2 because of our choice here to restrict the tilt angles to [-90$^{\circ}$, 90$^{\circ}$].  The uncertainties on the slopes increase with decreasing magnetic field, as the flux tube is more susceptible to deformation by convective flows, and the tilt angles exhibit more of a scatter about the best-fit line.  From white-light sunspot-group data spanning Solar Cycles 15 -- 21, \citet{dasi_2010} find an empirical Joy's Law equation of slope $m_\mathrm{A}=0.26 \pm 0.05$ for Mount Wilson sunspot data, and $m_\mathrm{A}=0.28 \pm 0.06$ for Kodaikanal data.  The values we report in columns 2 and 3 of Table \ref{tbl:tilts_radheat} agree with \citet{dasi_2010} within the reported uncertainties for both cases AB and RD in the low (15\,--\,30 kG) and high (100 kG) magnetic-field-strength regimes.
   
\begin{table}
\caption{Joy's Law best-fit line slopes and $\sigma_\mathrm{fit}$ values for cases AB (Columns 2, 4, 6) and RD (Columns 3, 5, 7).  (Column 1) magnetic field strengths.  (Column 2, 3) Slopes $m_\mathrm{A}$ (unitless) of the best-fit line following Method 1, and (Column 4, 5) slopes $m_\mathrm{B}$ (units of degrees) of the best-fit line following Method 2.  The uncertainties reported in relation to the slopes are uncertainties in the determination of the fit parameters $m_\mathrm{A}$ and $m_\mathrm{B}$, which takes into account both propagation of error and the variance of the data around the best-fit line.  Slopes of the best-fit lines following Methods 1 and 2 show a similar trend, reaching a maximum at 40 kG for case AB flux tubes, and 60 kG case RD flux tubes.  (Column 6, 7) Standard deviation $\sigma_\mathrm{fit}$ (units of degrees) of the tilt angle about the best-fit Joy's Law equation (Method 1).  For magnetic field strengths of $40-60$ kG, the values of $\sigma_\mathrm{fit}$ are substantially larger for case RD as compared to AB.}
\begin{tabular}{ccccccc}
\\
\hline
\emph{B} [kG] 	& $m_\mathrm{A-AB}$ 	& $m_\mathrm{A-RD}$	& $m_\mathrm{B-AB}$ 	& $m_\mathrm{B-RD}$	& $\sigma_\mathrm{AB}$				& $\sigma_\mathrm{RD}$\\
\hline	
100 	&	0.25 $\pm$ 0.02	&	0.32 $\pm$ 0.02	&	15.5$^{\circ}$ $\pm$ 1.1$^{\circ}$		&	19.8$^{\circ}$ $\pm$ 1.0$^{\circ}$			&	8.2$^{\circ}$						&	7.6$^{\circ}$\\

80 	&	0.29 $\pm$ 0.02	&	0.40 $\pm$ 0.02	&	17.7$^{\circ}$ $\pm$ 1.2$^{\circ}$		&	25.7$^{\circ}$ $\pm$ 1.2$^{\circ}$	 		&	9.1$^{\circ}$						&	10$^{\circ}$\\

60 	&	0.35 $\pm$ 0.02	&	0.44 $\pm$ 0.04	&	21.6$^{\circ}$ $\pm$ 1.5$^{\circ}$ 		&	27.1$^{\circ}$ $\pm$ 2.4$^{\circ}$			&	13$^{\circ}$						&	24$^{\circ}$\\

50 	&	0.38 $\pm$ 0.03	&	0.42 $\pm$ 0.04	&	23.8$^{\circ}$ $\pm$ 2.1$^{\circ}$ 		&	26.2$^{\circ}$ $\pm$ 2.6$^{\circ}$			&	18$^{\circ}$						&	26$^{\circ}$\\

40 	&	0.40 $\pm$ 0.04	&	0.27 $\pm$ 0.05	&	25.2$^{\circ}$ $\pm$ 2.5$^{\circ}$		&	16.7$^{\circ}$ $\pm$ 3.4$^{\circ}$			&	22$^{\circ}$						&	35$^{\circ}$\\

30 	&	0.31 $\pm$ 0.06	&	0.32 $\pm$ 0.05	&	19.6$^{\circ}$ $\pm$ 3.5$^{\circ}$		&	20.0$^{\circ}$ $\pm$ 3.3$^{\circ}$			&	33$^{\circ}$						&	34$^{\circ}$\\

15 	&	0.25 $\pm$ 0.07	&	0.31 $\pm$ 0.07	&	15.4$^{\circ}$ $\pm$ 4.2$^{\circ}$		&	19.1$^{\circ}$ $\pm$ 4.3$^{\circ}$	 		&	39$^{\circ}$						&	43$^{\circ}$\\
\hline
\end{tabular}
\label{tbl:tilts_radheat}
\end{table}
   
A Joy's Law fit can also be performed using the equation (Method 2) $\alpha=m_\mathrm{B} \sin \theta$,
which is a good choice assuming the origin of the tilt angle is related to the Coriolis force, as this force varies with latitude as sin$(\theta)$.  Here, the tilt angle $\alpha$ is again in units of degrees and $m_\mathrm{B}$ also has units of degrees.  The values we obtain from our simulations are shown in columns 4 and 5 of Table \ref{tbl:tilts_radheat}.  \citet{stenflo_2012} perform such a fit using 15 years of MDI full-disk magnetograms, finding a slope of $m_\mathrm{B}=32.1^{\circ}$ $\pm$ $0.7^{\circ}$.  However, using Mount Wilson sunspot-group data from 1917\,--\,1985, \citet{fisher_1995} find a best-fit equation slope of $m_\mathrm{B}$=$15.69^{\circ}$ $\pm$ $0.66^{\circ}$.   This large difference in $m_\mathrm{B}$ obtained from observations might occur because weaker active regions, which may not appear in white-light sunspot-group images, can be identified in magnetograms.  However, it may also be the result of selection effects employed by the authors.  For both case AB and case RD flux tubes, the values that we obtain for $m_\mathrm{B}$ are too small compared to the magnetogram data of \citet{stenflo_2012}, but too large for most magnetic field strengths compared to the sunspot-group data of \citet{fisher_1995}.  

The tilt-angle slopes (both $m_\mathrm{A}$ and $m_\mathrm{B}$, as their behaviors mirror each other) for case RD flux tubes peak at 60 kG, whereas they peak at 40 kG for case AB flux tubes of $10^{22}$ Mx.  In both cases, this peak corresponds to the magnetic field strength where the flux tube takes the longest time to emerge (see Figure \ref{fig:risetimes_rad_adia}).  This behavior can be understood briefly as follows (see also Article 1).  A larger average emergence time occurs when the effects due to magnetic buoyancy and the average convective downflows are of similar magnitudes, \textit{i.e.} when the two sides of Equation (\ref{eq:compare2}) are approximately equal.  Figure \ref{fig:compare_alf_adia_rad} illustrates this point well for a case AB 40 kG flux tube and a case RD 60 kG flux tube.  Since these flux tubes spend a longer time on average in the convection zone proper, the joint effects of the Coriolis force and the average kinetic helicity of the convection simulation on the rising flux tube apex help to boost the tilt angle of the flux tube toward the equator, increasing the Joy's Law trend.  

In comparison to the slopes $m_\mathrm{A}$ and $m_\mathrm{B}$ for case AB flux tubes of $10^{22}$ Mx, the slopes of the best-fit lines that we obtain for case RD flux tubes are larger for flux tubes of $60-100$ kG, and smaller for flux tubes of 40 kG.  These differences are statistically significant (\textit{i.e.} larger than the uncertainties).  We expected that the overall tilt-angle best-fit slope would be reduced for case RD flux tubes because the value of $\delta T = T_\mathrm{e}-T$ is decreased, initiating a converging parallel flow of mass elements at the flux tube apex at greater depths \citep[\textit{e.g.}][]{fan_1994}.  Due to this converging flow, the Coriolis force acts to tilt the flux tube apex away from the Equator in either hemisphere (opposite the Joy's Law trend), thereby reducing the tilt angle.  

We suggest that the increased tilt-angle trend of $60-100$ kG case RD flux tubes, as compared to case AB flux tubes, is related to a decrease in magnetic field strength at the flux tube apex, and that subsequently the flux tube becomes more susceptible to the the effects of helical convective upflows.  This is capable of overcoming the Coriolis force induced tilt of the wrong sense (\textit{i.e.} away from the Equator) due to a converging parallel flow at the flux tube apex.  However, this does not explain why case RD 40 kG flux tubes show a reduced tilt-angle trend compared to case AB flux tubes.  While case RD flux tubes of $\le$60 kG do not anchor, one or both of the footpoints of the buoyantly rising loops are usually located within the lower third of the convection zone (see Figure \ref{fig:radial_vr_stuff_rad}).  However, for case RD flux tubes of 40 kG, the flux tube as a whole resides in the middle of the convection zone throughout much of its evolution, where it is continually pummeled by convective upflows and downflows.  Buoyant loops of these flux tubes that reach the simulation upper boundary tend to have their footpoints located in the middle of the convection zone (see again Figure \ref{fig:radial_vr_stuff_rad}).  Perhaps the fact that in this regime the flux tube footpoints are higher in the convection zone also contributes to the reduced tilt angle trend.  If stretching of the flux tube loop contributes to the tilting motion supplied by the Coriolis force \citep[\textit{e.g.}][]{dsilva_1993}, then loops with troughs that do not extend to the base of the convection zone will experience a reduced tilt angle.  The combined effects of the reduced stretching of the loop and the increased converging parallel flow at the flux tube apex due to a reduction in $\delta T$ may be strong enough at 40 kG magnetic field strengths to overcome the increased coupling to convective helical upflows in the non-adiabatic regime.  

\subsubsection{Tilt Angle Scatter About the Joy's Law Trend}
\label{sec:scatter_joy}

To quantify the scatter of the tilt angles around the best-fit line, a phenomenon introduced by convection, we calculate the standard deviation of the tilt about its fitted value following:
\begin{equation}
\sigma_\mathrm{fit}=\sqrt{\frac{ \sum_{i=1}^{N} (\alpha_{i}-\alpha_\mathrm{fit})^{2}}{N}},
\label{eqn:scatter}
\end{equation} 
where $\alpha_{i}$ is the $i$th tilt angle, $\alpha_\mathrm{fit}$ is the $i$th tilt angle as a result of the fit following Method 1, $N$ is the number of points considered, and $\alpha$ is in units of degrees.  We evaluate $\sigma_\mathrm{fit}$ for each field strength, where the results are given in columns 6 and 7 of Table \ref{tbl:tilts_radheat}.  When calculating $\sigma_\mathrm{fit}$, we have shifted the tilt angles appropriately such that all tilts angles fall within the range [-90$^{\circ}$, $90^{\circ}$].  This is unlike what was done in Article 2 where tilt angles were considered to fall within the range of [-$180^{\circ}$, $180^{\circ}$], thereby retaining anti-Hale tilt-angle information.  Due to the tilt-angle scatter introduced by the increased sensitivity of case RD flux tubes to convection, the values of $\sigma_\mathrm{fit}$ for the $40-60$ kG case RD flux tubes are larger than case AB flux tubes of the same magnetic field strength.  Using Mount Wilson white-light sunspot-group data, \citet{fisher_1995} found that $\sigma_\mathrm{fit}\approx30^{\circ}$.  A value of $\sigma_\mathrm{fit}\le30^{\circ}$ is found for case RD flux tubes of $\ge$50 kG, and $\ge$40 kG for case AB flux tubes.  Similar values of $\sigma_\mathrm{fit}\le30^{\circ}$ are obtained when the fit following Method 2 is used instead.   

\subsubsection{Preferred Tilt Angle}       
\label{sec:preferred}

\begin{figure}
\centerline{\includegraphics[scale=.55]{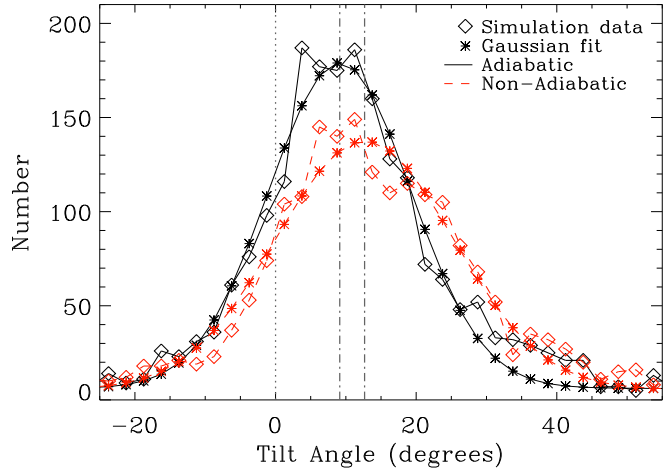}}
\caption{Distribution of tilt angles in 2.5$^{\circ}$ bins, shown for both the case AB (adiabatic, black solid lines) and case RD (radiative diffusion, red dashed lines) simulations.  For each simulation type, the simulation data is distinguished by diamond symbols, and the Gaussian fit to the data is distinguished by asterisks.  The dotted line shows the zero mark along the horizontal axis, while the dash-dotted lines show the centers of the Gaussian fits, which are $12.7^{\circ}$ for the case RD simulations, and $9.1^{\circ}$ for the case AB simulations.  The addition of radiative diffusion to the simulation model increases the preferred tilt angle, but decreases the peak of the distribution.}
\label{fig:histogram}
\end{figure}

The distribution of tilt angles from our $10^{22}$ Mx flux tube simulations are shown in Figure \ref{fig:histogram}, where we compare both case AB and case RD flux tubes.  Unlike the previous analyses in Section \ref{sec:tilt_rad}, we allow the tilt angles to fall within the [-$180^{\circ}$, $180^{\circ}$] range: however, we only plot the range between [-$25^{\circ}$, $55^{\circ}$] so as to highlight the difference in the two distributions near their centers.  A non-linear least squares Gaussian fit to the distribution gives a center, or preferred tilt angle, of $12.7^{\circ}\pm3.6^{\circ}$ for case RD flux tubes, and $9.1^{\circ}\pm3.2^{\circ}$ for case AB flux tubes.  The uncertainty on the preferred tilt angle is the standard deviation of the Gaussian fit.  We note an increase in the center of the Gaussian fit for flux tubes with the inclusion of radiative heating, and also a wider distribution around the center.  This increase in tilt angle reflects the results found in Section \ref{sec:joy_law}.  However, given the large standard deviation of the Gaussian fits, the centers of the distributions are not statistically different.  Using a similar analysis, \citet{howard_1996} found a tilt-angle distribution that peaks between 2.5$^{\circ}$\,--\,5$^{\circ}$ for sunspot-group tilt angles derived from Mount Wilson white-light photographs (1917\,--\,1985), and between 7.5$^{\circ}$\,--\,10$^{\circ}$ for sunspot groups derived from plage in Mount Wilson daily magnetograms (1967\,--\,1995).  

Although both the case AB and case RD simulations each are comprised of $\approx$2300 simulations, we note that the peak (\textit{i.e.} height) of the distribution for case AB flux tubes is larger than for the RD case.  There are two reasons for this behavior.  First, the case RD distribution is broader than for case AB.  Secondly, and most importantly, the case RD simulations have a larger percentage of tilt angles that are anti-Hale ($\ge |90^{\circ}|$); $9.9\,\%$ as compared to $5.0\,\%$ for the case AB simulations.  In comparison, $\approx$$4\,\%$ of medium to large-sized active regions exhibit anti-Hale tilt angles \citep{wang_1989,stenflo_2012}.  So that the case RD simulations exhibit only $\approx$$4\,\%$ anti-Hale tilt angles, we would have to exclude flux tubes with magnetic field strengths of $\le$40 kG.  In this case, the center of the tilt-angle distribution drops to $11.5^{\circ}\pm3.3^{\circ}$.  

\section{A Note on Flux Tube Storage in the Convective Overshoot Region}
\label{sec:flux_storage}

In one solar dynamo paradigm, it is suggested that the dynamo generated magnetic field is stored in a region of overshooting convective motions (overshoot region) at the base of the convection zone until becoming buoyant enough to rise toward the surface \citep[\textit{e.g.}][]{spiegel_1980,galloway_1981,miesch_lrsp}.  It has been recognized that while the sub-adiabatic stratification of the overshoot region stabilizes magnetic flux tubes against buoyancy instabilities, the inflow of heat to the flux tube due to radiative diffusion forces the flux tube to rise quasi-statically (\textit{i.e.} moving through a sequence of equilibria such that all forces closely balance) out of the overshoot region in $\approx$1 year or less, short compared to the $\approx$11 year solar cycle period \citep[\textit{e.g.}][]{vanballe_1982,fan_1996,rempel_2003}.  In this section, we review the nature of the solar overshoot region (Section \ref{sec:solar_overshoot}), and examine the upward drift of flux tubes in the overshoot region due to radiative diffusion (Section \ref{sec:drift}).

\subsection{The Nature of the Solar Convective Overshoot Region}
\label{sec:solar_overshoot}
Generally, the base of the convection zone is defined as the radius where the stratification changes from nearly adiabatic ($\nabla_\mathrm{e}\approx\nabla_\mathrm{ad}$) to substantially sub-adiabatic ($\nabla_\mathrm{e}<\nabla_\mathrm{ad}$) \citep[\textit{e.g.}][]{miesch_lrsp}.  Information about the sound-speed gradient in the solar interior, related to the temperature gradient, can be obtained from inversions of helioseismic data, revealing that the base of the convection zone occurs at $0.713\pm0.003$\,R$_{\odot}$ \citep{jcd_1991}.  In reality, this observationally derived extent of the convection zone may include near-adiabatic portions of the overshoot region that are weakly sub-adiabatic (\textit{i.e.} $\delta=\nabla_\mathrm{e}-\nabla_\mathrm{ad}$ is negative, but $\nabla_\mathrm{e}$ is still very close to $\nabla_\mathrm{ad}$).  Helioseismology has also been used to identify the tachocline at a radius of $0.693\pm0.003$\,R$_{\odot}$ near the equator \citep{char_1999,basu_2003}, indicating that the tachocline lies below the base of the convection zone, although it may still be located within the overshoot region.  

Convective plumes will penetrate some distance into the stably stratified region below the base of the convection zone before buoyancy decelerates them, influencing the \emph{stiffness} of the transition from super-adiabatic to sub-adiabatic temperature gradients and the thickness of the overshooting layer.  As suggested by \citet{zahn_1991}, efficient overshooting convection renders the temperature gradient weakly sub-adiabatic.  Only when the convective motions have been decelerated to the point where turbulent mixing becomes inefficient compared to thermal diffusion will the temperature gradient adjust over some distance to that of the radiative interior.  

Much uncertainty still exists regarding the depth of the overshoot region and its mean sub-adiabatic temperature gradient.  Helioseismic inversions find an upper limit on the thickness of $\approx$$0.05H_\mathrm{p}$, less than 0.01\,R$_{\odot}$  \citep{basu_1994,basu_1997,monteiro_1994}.  Early attempts to model the overshoot region predicted thicknesses of $0.2\,H_\mathrm{p}$ to $0.4\,H_\mathrm{p}$, weakly sub-adiabatic stratifications of $\delta\approx-10^{-6}$ to $-10^{-5}$, and a sharp transition in the temperature gradient between the nearly adiabatic overshoot region and the radiative interior underneath \citep[\textit{e.g.}][]{vanballe_1982,schmitt_1984,pidatella_1986,skaley_1991}.  However, such small values of $\delta$ are not sufficient enough to facilitate the storage of magnetic flux on solar-cycle timescales (see Section \ref{sec:drift}).  It may be the case that strong magnetic fields in the overshoot region can suppress penetrative convective motions, reducing the convective energy transport and convective heat conductivity, leading to a stronger sub-adiabaticity of $\delta\le-10^{-4}$ \citep{rempel_2003}.

Numerical convection simulations with overshoot by \citet{brummell_2002} exhibit a strong sub-adiabatic overshoot layer with a smoother transition in the temperature gradient toward the radiative value, which may be more inline with the actual conditions in the Sun \citep[\textit{e.g.}][]{jcd_2011}.  This model also produces larger overshooting penetration depths of $0.4-2\,H_\mathrm{p}$.  \citet{rempel_2004} points out that the strongly sub-adiabatic overshoot found by numerical simulations is likely due to the larger thermal diffusivities required by numerical constraints, resulting in a larger energy flux.  The efficiency of the mixing between convective upflows and downflows, as well as their velocities, may determine the depth of the overshoot region and the temperature gradient transition from nearly adiabatic to that of the radiative interior.

For the purposes of our simulations, we have identified the base of the convection zone from our reference stellar structure model (Model S) as $r_\mathrm{czb}=0.723$\,R$_{\odot}$, the radius where the stratification changes from super-adiabatic ($\delta>0$) to sub-adiabatic ($\delta<0$).  We model a simple convective overshoot region where $\delta$ decreases from 0 to $-10^{-3}$ over a distance of $\approx10^{9}$ cm (0.01\,R$_{\odot}$) below $r_\mathrm{czb}$, then remains fixed at $-10^{-3}$ to the bottom of our computation domain at 0.69\,R$_{\odot}$.  This value of $\delta$ was chosen because it is found in Section \ref{sec:drift} that $\delta$ in the range of $-10^{-3}$ to $-10^{-4}$ is necessary for active-region-scale magnetic flux tubes to remain stored in the overshoot region for solar-cycle timescales.  We do not include a transition from the sub-adiabatic stratification of the overshoot region to the temperature gradient of the radiative interior, as our flux tubes never reach such depths.

\subsection{Upward Drift of Flux Tubes in the Convective Overshoot Region}
\label{sec:drift}

Flux tubes initially in a neutrally buoyant state evolve very differently in the sub-adiabatic overshoot region compared to the super-adiabatic convection zone as discussed in Sections \ref{sec:dynamics_rad}\,--\,\ref{sec:emergence_properties}.  In the overshoot region, the buoyancy of the flux tube will be suppressed as the adiabatic component of the flux tube's buoyancy [$\mathrm{d}\Delta \rho/\mathrm{d}t$] due to adiabatic expansion of the rising flux tube (Equation (\ref{eq:drho_ad})) takes on a negative value because of the sub-adiabatic temperature gradient (\textit{i.e.} $\delta<0$).  A flux tube will rise quasi-statically through the overshoot region if the thermal timescale over which radiative heating significantly increases the flux tube's buoyancy from a neutrally buoyant state ($\tau_\mathrm{th}\approx10^{7}$ seconds) is long compared to the dynamic timescale characterized by the magnetic Brunt-V\"ais\"al\"a frequency ($\tau_\mathrm{B}\approx10^{5}$ seconds) \citep{fan_1996,fan_2009}.  This means that the flux tube will evolve through a series of mechanical equilibrium states during which the buoyancy remains close to zero.

\citet{fan_1996} derived an expression for the quasi-static rise velocity $v_\mathrm{r}$ (rise speed) of the flux tube in the overshoot region, and subsequently a characteristic rise time for an upward drift of the flux tube out of the overshoot region.  For simplicity, they consider a uniform horizontal flux tube that remains nearly neutrally buoyant (\textit{i.e.} in quasi-equilibrium) in a plane-parallel overshoot region.  Under these assumptions, the upward drift of the flux tube is found to be:
\begin{equation}
v_\mathrm{r} = \frac{H_\mathrm{p}}{p_\mathrm{e}} \nabla_\mathrm{ad} \biggl( \frac{\mathrm{d}Q}{\mathrm{d}t} \biggr)_{1} \biggl[-\delta +\frac{1}{\beta} \biggl( \frac{2}{\gamma^{2}} -\frac{1}{\gamma} \biggr) \biggr]^{-1},
\end{equation}  
where $\delta=\nabla_\mathrm{e}-\nabla_\mathrm{ad}$, and is referred to as the sub-adiabaticity in a stably stratified region where $\delta$ takes on a negative value, and $1/\beta=B^{2}/8\pi p_\mathrm{e}$.  For typical values in the overshoot region, $H_\mathrm{p}\approx$ 6\e{9} cm, $p_\mathrm{e}\approx$ 6\e{13} g cm$^{-1}$ s$^{-2}$, $(\mathrm{d}Q/\mathrm{d}t)_{1}\approx30$ erg cm$^{-3}$ s$^{-1}$, $\nabla_\mathrm{ad}\approx0.4$, and $\gamma\approx5/3$.  Then an estimate for the rise speed [cm s$^{-1}$] of the flux tube in the overshoot region becomes:
\begin{equation}
v_{r}\approx 1.2 \times 10^{-3} \biggl[ -\delta + \frac{0.12}{\beta} \biggr]^{-1}.
\end{equation}  
 
\begin{table}
\caption{The rise speed $v_\mathrm{r}$ (Column 2) and rise time $\tau$ (Column 3) for a 100 kG, $10^{22}$ Mx flux tube to escape from the overshoot region for varying estimates of the super-adiabaticity $\delta$ (Column 1).  In comparison, the length of the solar cycle is only $\approx$11 years.}
\begin{tabular}{ccc}
\\
\hline
$\delta$ & $v_\mathrm{r}$ [cm s$^{-1}$] & $\tau$ \\
\hline	
$-10^{-3}$		&	1.2	&	26 years \\
$-10^{-4}$		&	12	&	2.6 years \\
$-10^{-5}$		&	110	&	98 days \\
$-10^{-6}$		&	670	&	17 days \\
\hline
\end{tabular} 
\label{tbl:rise_speeds}
\end{table} 
 
In the overshoot region, $\beta$ ranges from 1.5\e{5} for 100 kG magnetic fields to 6.7\e{6} for 15 kG magnetic fields.  The rise speed of the flux tube is highly dependent on the super-adiabaticity [$\delta$] of the overshoot region, which is most likely in the range of $-10^{-3}$ to $-10^{-6}$ \citep[\textit{e.g.}][]{rempel_2003,rempel_2004}.  As $\delta$ approaches zero, the term $0.12/\beta$ becomes increasingly important.  The quantity $0.12/\beta$ is largest for larger magnetic field strengths; thus 100 kG flux tubes have a slower upward drift out of the overshoot region.  Using $B_{0}=100$ kG, and for different estimates of the value $\delta$, the rise speed $v_\mathrm{r}$ and time $\tau$ for a flux tube to escape from the overshoot region are shown in Table \ref{tbl:rise_speeds}.  Here we have assumed the depth of the overshoot region to be $10^{9}$ cm and that $\delta$ is constant throughout this layer.  For flux tubes to remain stored in the overshoot region for the length of the solar cycle, $\delta$ needs to be somewhere between $-10^{-4}$ and $-10^{-3}$.  This is in agreement with \citet{moreno_1992}, where it is found that $\delta\le-10^{-5}$ for $B_{0}\le100$ kG flux tubes to be stored in the overshoot region.  The thickness of the overshoot layer that we have assumed is slightly thicker than estimates derived from helioseismic inversions, but shallower than what most solar models predict.  A thicker overshoot region would certainly result in longer escape times.  Rather than assuming an overshoot region with constant $\delta$, a spatially varying $\delta$ that increases from an average value to 0 upward from the base of the overshoot region would likely decrease the escape time.

Some work suggests that convective flows may be able to help facilitate magnetic flux storage, rather than relying on a strong sub-adiabaticity of $\delta\le -10^{-4}$ to store flux tubes in the overshoot region for the length of the solar cycle. Strong downflow plumes in the lower convection zone could effectively pump magnetic flux into the overshoot region, and may also help to keep flux from escaping \citep[\textit{e.g.}][]{tobias_2001,brummell_2002,browning_2006}.  Or, perhaps an equatorward meridional flow present in the overshoot region may be sufficient to keep flux tubes stably stored for a solar cycle \citep{vanballe_1982,vanballe_1988}.  However, \citet{rempel_2003} found that such a scenario is only possible for thin flux tubes containing less than $1\,\%$ of a typical sunspot magnetic flux, well below the magnetic flux considered in this article.

It is as of yet not clear whether magnetic flux tubes are generated at a tachocline interface in the overshoot region of the Sun; however, this is the paradigm that we adopt for this study.  For the thin flux tube simulations used in this work and Articles 1 and 2, it is assumed that the flux tube has already risen out of the overshoot region.  Flux tubes with footpoints that do anchor in the overshoot region do so because convective motions initiate buoyancy instabilities, which amplify undulations, causing the flux tube troughs to penetrate into the stably stratified plasma.  In Articles 1 and 2 where radiative diffusion is not considered, all flux tubes that evolve with convection have footpoints that penetrate into the overshoot region.  However, as described in Section \ref{sec:morph},  when radiative heating is considered, flux tubes of $\le$60 kG do not anchor in the overshoot region, while flux tubes of $\ge$80 kG do develop anchored footpoints that continue to drift out of the overshoot region as the flux tube evolves in time.  

\begin{figure}
\centerline{\includegraphics[scale=.6]{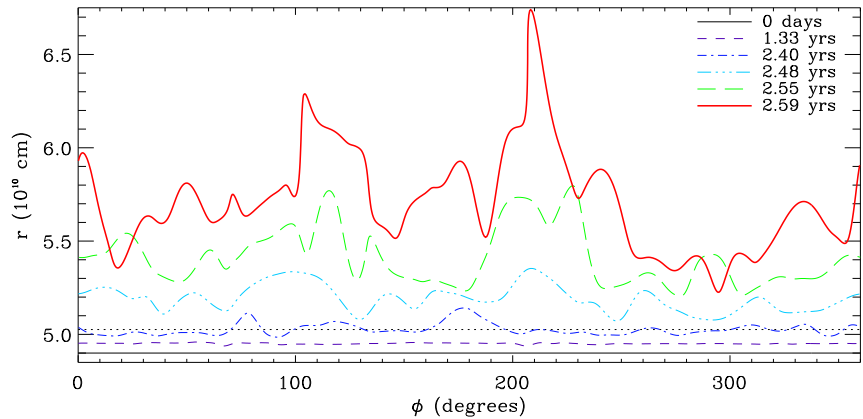}}
\caption{Representative example of a $B_{0}=60$ kG case RD flux tube as it rises quasi-statically through the sub-adiabatic overshoot region and drifts upward into the super-adiabatic convection zone, with the base of the convection zone ($r_\mathrm{czb}$) represented by the dotted line.  The $\theta_{0}=12^{\circ}$ flux tube is initiated below $r_\mathrm{czb}$ at 4.9\e{10} cm, and the value of $\delta$ in the overshoot region decreases downward from 0 at $r_\mathrm{czb}$ to $-10^{-4}$ over a distance of $\approx10^{9}$ cm.  The flux tube drifts upward through the overshoot region in $\approx2.4$ years, long compared to the $0.2$ years it takes for the flux tube to traverse the convection zone.}
\label{fig:overshoot_tube}
\end{figure}

We have performed some simulations where the flux tube is initiated in the stable overshoot region and allowed to rise quasi-statically under the influence of radiative heating and overshooting convection.  We find that the flux tube drifts out of the overshoot region more or less as a whole regardless of (i) the initial field strength of the flux tube, (ii) initial depth below $r_\mathrm{czb}$, (iii) average sub-adiabaticity of the overshoot region , or (iv) steepness of the transition from $\delta=0$ to that of the average sub-adiabaticity of the overshoot region.  Once the toroidal flux tube has emerged as a whole from the overshoot region, the evolution of the flux tube proceeds in a similar way as when we initiate the flux tube 2.4 Mm above the base of the convection zone, as is the procedure following the discussion in Section \ref{sec:model}.   In order to initiate the flux tube in the overshoot region such that it always remains anchored except for buoyantly rising loops, we would have to artificially introduce an entropy perturbation to the flux tube of large amplitude, as was done by \citet{fan_1996}, rather than allowing convective flows to provide the perturbations self-consistently.

Figure \ref{fig:overshoot_tube} shows the evolution of a $B_{0}=60$ kG, $\theta_{0}=12^{\circ}$ case RD flux tube that is initially placed in the overshoot region at 4.9\e{10} cm, $\approx$$10^{9}$ cm below $r_\mathrm{czb}$.  For this particular simulation, $\delta$ decreases from 0 to $-10^{-4}$ over a distance of $\approx$$10^{9}$ cm.  The flux tube takes $\approx$$2.4$ years to rise through the overshoot region, in agreement with the values quoted in Table \ref{tbl:rise_speeds}.  Once it emerges from the overshoot region the flux tube does not anchor, rising toward the surface in a little under two months (0.2 years).  At the top of our computation domain (0.97\,R$_{\odot}$), the flux tube apex emerges at a latitude of $\theta_\mathrm{em}=31.2^{\circ}$ with a tilt angle of $-10.1^{\circ}$, the appropriate sign for the northern hemisphere.  All of these properties are consistent with the 60 kG case RD flux tubes discussed in Sections \ref{sec:dynamics_rad} and \ref{sec:emergence_properties}.

\section{Summary and Discussion}
\label{sec:discuss_rad}

In this article, we modify the energy equation of our thin flux tube model to include the effect of flux tube heating due to radiative diffusion, unlike our simulations in Articles 1 and 2 where the flux tube is assumed to evolve adiabatically.  This allows us to study the influence of radiative diffusion, in conjunction with solar-like convective flows, on the dynamic evolution of active-region-scale magnetic flux tubes.  

As a result of the inclusion of radiative diffusion to the thin flux tube energy equation, we find that flux tubes of $\le$60 kG subject to convective flows are no longer able to anchor in the overshoot region, unlike the flux tubes discussed in Articles 1 and 2.  Heating of the flux tubes supplied by radiative diffusion especially in the lower convection zone significantly increases the buoyancy of the flux tube earlier in its evolution.  Flux tubes of $\le$60 kG now float away from the base of the convection zone before magnetic buoyancy instabilities can set in to anchor the troughs of the flux tube in the overshoot region.  This uniformly increased buoyancy of the flux tube early in its evolution results in a rise time of $\le$0.2 years for all flux tubes, which is significantly less than the maximum rise times of adiabatically evolving flux tubes by $\approx$0.5 years.   

Unlike adiabatically evolving flux tubes of $10^{22}$ Mx, flux tubes of $\le$60 kG that evolve with the addition of radiative heating exhibit a larger poleward deflection, especially at mid-to-high latitudes of $\approx$$15^{\circ}$\,--\,$40^{\circ}$.  The drift of $\le$60 kG flux tubes from the base of the convection zone and subsequent lack of anchoring facilitates a poleward slippage, which is partially mitigated by convective flows.  When radiative heating is included in the thin flux tube energy equation, we also find that all magnetic-field-strength flux tubes of $10^{22}$ Mx still exhibit a Joy's Law trend in agreement with observations.  However, flux tubes with magnetic field strengths of $\le$40 kG exhibit too large a tilt-angle scatter and too large a percentage of anti-Hale tilt angles as compared to solar observations.  These results point toward larger magnetic field strengths of $\ge$50 kG as the progenitors of solar active regions, inline with the adiabatically evolving flux tube simulations of Articles 1 and 2.

We also discuss the problem of flux storage in the overshoot region when the effect of radiative diffusion on flux tube evolution is considered, suggesting that the time for flux tubes to escape from the overshoot region is highly sensitive to the value of the super-adiabaticity [$\delta$].  When we allow flux tubes to originate in the overshoot region, all magnetic-field-strength flux tubes rise quasi-statically through the overshoot region.  Once they drift out of the overshoot region as a whole, the evolution of the flux tube proceeds in a similar way as when we initiate the flux tube 2.4 Mm above the base of the convection zone.

In our treatment of the heat input per unit volume of the flux tube [$\mathrm{d}Q_\mathrm{v}/\mathrm{d}t$], we only include $(\mathrm{d}Q/\mathrm{d}t)_{1}$, the contribution from the deviation of the mean temperature gradient in the external plasma environment from that of radiative equilibrium.  We have here neglected $(\mathrm{d}Q/\mathrm{d}t)_{2}$, the contribution from radiative diffusion across the flux tube due to the temperature difference between the flux tube and the external plasma environment.  This is only a reasonable approximation for flux tubes in the range of $15-100$ kG for magnetic flux values of $10^{22}$ Mx, on the order of the largest scale active regions.  If $(\mathrm{d}Q/\mathrm{d}t)_{2}$ were included and we performed simulations for flux tubes of $10^{20}-10^{21}$ Mx, we suspect that the average rise times for these simulations would be slightly less than those reported in Section \ref{sec:rise_times}.  Following the results of Articles 1 and 2 concerning tilt angle trends, we expect for smaller values of magnetic flux that the scatter of the tilt angle about the Joy's Law trend will increase with decreasing flux, as will the percentage of anti-Hale tilt angles.      

The simulations performed in this article lead us to a better and more comprehensive understanding of the processes involved in flux emergence on the Sun and solar-like stars.  It is likely that if flux tubes are generated in the overshoot region by the dynamo process, they drift upward out of the overshoot region due to heating of the tube plasma as a result of the deviation of the temperature gradient from that of radiative equilibrium.  Once out of the overshoot region, flux tubes then drift away from the base of the convection zone as a whole due to an enhanced buoyancy supplied by radiative heating, only anchoring in the overshoot region if undular magnetic buoyancy instabilities grow sufficiently fast.  Only when reaching the middle to upper convection zone is the dynamic evolution of the flux tube well approximated by adiabatic evolution.  Our simulations show that flux tubes do not necessarily need to have anchored footpoints in the overshoot region to produce emergence properties similar to those of active regions on the Sun.

%

%
\begin{acks}
The majority of this work was conducted while M.A. Weber was a Graduate Research Fellow at High Altitude Observatory (HAO), a division of the National Center for Atmospheric Research (NCAR).  NCAR is sponsored by the National Science Foundation.  This research was sponsored in part by NASA SHP grant NNX10AB81G to NCAR.  M.A. Weber is now supported by the European Research Council under grant agreement 337705 (the CHASM project).  We would like to thank Mark Miesch (HAO/NCAR) for providing the ASH convective flows used in this article.  The authors are also grateful to Mausumi Dikpati (HAO/NCAR) and the anonymous referee for reading our manuscript and providing helpful comments.
\end{acks}

\section*{Disclosure of Potential Conflicts of Interest}
The authors declare that they have no conflicts of interest.

%
%

\bibliographystyle{spr-mp-sola}

\begin{thebibliography}{59}
\ifx\bisbn     \undefined \def\bisbn  #1{ISBN #1}\fi
\ifx\binits    \undefined \def\binits#1{#1}\fi
\ifx\bauthor   \undefined \def\bauthor#1{#1}\fi
\ifx\batitle   \undefined \def\batitle#1{#1}\fi
\ifx\bjtitle   \undefined \def\bjtitle#1{\textit{#1}}\fi
\ifx\bvolume   \undefined \def\bvolume#1{\textbf{#1}}\fi
\ifx\byear     \undefined \def\byear#1{#1}\fi
\ifx\bissue    \undefined \def\bissue#1{#1}\fi
\ifx\bfpage    \undefined \def\bfpage#1{#1}\fi
\ifx\blpage    \undefined \def\blpage #1{#1}\fi
\ifx\burl      \undefined \def\burl#1{\textsf{#1}}\fi
\ifx\href      \undefined \def\href#1#2{\textsf{#2}}\fi
\ifx\betal     \undefined \def\betal{\textit{et al.}}\fi
\ifx\bctitle   \undefined \def\bctitle#1{#1}\fi
\ifx\beditor   \undefined \def\beditor#1{#1}\fi
\ifx\bbtitle   \undefined \def\bbtitle#1{\textit{#1}}\fi
\ifx\bedition  \undefined \def\bedition#1{#1}\fi
\ifx\bseriesno \undefined \def\bseriesno#1{\textbf{#1}}\fi
\ifx\blocation \undefined \def\blocation#1{#1}\fi
\ifx\bsertitle \undefined \def\bsertitle#1{\textit{#1}}\fi
\ifx\bsnm      \undefined \def\bsnm#1{#1}\fi
\ifx\bsuffix   \undefined \def\bsuffix#1{#1}\fi
\ifx\bparticle \undefined \def\bparticle#1{#1}\fi
\ifx\barticle  \undefined \def\barticle#1{}\fi
\ifx\binstitute  \undefined \def\binstitute#1{#1}\fi
\ifx\bpublisher  \undefined \def\bpublisher#1{#1}\fi
\ifx\doiurl    \undefined
  \def\doiurl#1{\href{http://dx.doi.org/#1}{\textsf{DOI}}}\fi
\ifx\arxivurl  \undefined
  \def\arxivurl#1{\href{http://arxiv.org/abs/#1}{\textsf{arXiv}}}\fi
\ifx\adsurl    \undefined
  \def\adsurl#1{\href{http://adsabs.harvard.edu/abs/#1}{\textsf{ADS}}}\fi
\ifx\botherref \undefined \def\botherref#1{}\fi
\ifx\url       \undefined \def\url#1{\textsf{#1}}\fi
\ifx\bchapter  \undefined \def\bchapter#1{}\fi
\ifx\bbook     \undefined \def\bbook#1{}\fi
\ifx\bcomment  \undefined \def\bcomment#1{#1}\fi
\ifx\oauthor   \undefined \def\oauthor#1{#1}\fi
\ifx\citeauthoryear \undefined\def \citeauthoryear#1{#1}\fi
\def\endbibitem {}
\ifx\bconflocation  \undefined \def\bconflocation#1{#1} \fi

\bibitem[\protect\citeauthoryear{{Achterberg}}{1996}]{achterberg_1996}
\begin{barticle}
\bauthor{\bsnm{{Achterberg}}, \binits{A.}}:
\byear{1996},
\batitle{{Variational principle for slender flux tubes. I. General equations
  and added mass effects.}}
\bjtitle{\aap}
\bvolume{313},
\bfpage{1008}.
\adsurl{1996A\%26A...313.1008A}.
\end{barticle}
\endbibitem

\bibitem[\protect\citeauthoryear{{Basu}}{1997}]{basu_1997}
\begin{barticle}
\bauthor{\bsnm{{Basu}}, \binits{S.}}:
\byear{1997},
\batitle{{Seismology of the base of the solar convection zone}}.
\bjtitle{\mnras}
\bvolume{288},
\bfpage{572}.
\adsurl{1997MNRAS.288..572B}.
\end{barticle}
\endbibitem

\bibitem[\protect\citeauthoryear{{Basu} and {Antia}}{2003}]{basu_2003}
\begin{barticle}
\bauthor{\bsnm{{Basu}}, \binits{S.}},
\bauthor{\bsnm{{Antia}}, \binits{H.M.}}:
\byear{2003},
\batitle{{Changes in Solar Dynamics from 1995 to 2002}}.
\bjtitle{\apj}
\bvolume{585},
\bfpage{553}.
\doiurl{10.1086/346020}.
\adsurl{2003ApJ...585..553B}.
\end{barticle}
\endbibitem

\bibitem[\protect\citeauthoryear{{Basu}, {Antia}, and
  {Narasimha}}{1994}]{basu_1994}
\begin{barticle}
\bauthor{\bsnm{{Basu}}, \binits{S.}},
\bauthor{\bsnm{{Antia}}, \binits{H.M.}},
\bauthor{\bsnm{{Narasimha}}, \binits{D.}}:
\byear{1994},
\batitle{{Helioseismic measurement of the extent of overshoot below the solar
  convection zone}}.
\bjtitle{\mnras}
\bvolume{267},
\bfpage{209}.
\adsurl{1994MNRAS.267..209B}.
\end{barticle}
\endbibitem

\bibitem[\protect\citeauthoryear{{Browning}
  \textit{et~al.}}{2006}]{browning_2006}
\begin{barticle}
\bauthor{\bsnm{{Browning}}, \binits{M.K.}},
\bauthor{\bsnm{{Miesch}}, \binits{M.S.}},
\bauthor{\bsnm{{Brun}}, \binits{A.S.}},
\bauthor{\bsnm{{Toomre}}, \binits{J.}}:
\byear{2006},
\batitle{{Dynamo Action in the Solar Convection Zone and Tachocline: Pumping
  and Organization of Toroidal Fields}}.
\bjtitle{\apjl}
\bvolume{648},
\bfpage{L157}.
\doiurl{10.1086/507869}.
\adsurl{2006ApJ...648L.157B}.
\end{barticle}
\endbibitem

\bibitem[\protect\citeauthoryear{{Brummell}, {Clune}, and
  {Toomre}}{2002}]{brummell_2002}
\begin{barticle}
\bauthor{\bsnm{{Brummell}}, \binits{N.H.}},
\bauthor{\bsnm{{Clune}}, \binits{T.L.}},
\bauthor{\bsnm{{Toomre}}, \binits{J.}}:
\byear{2002},
\batitle{Penetration and overshooting in turbulent compressible convection}.
\bjtitle{\apj}
\bvolume{570},
\bfpage{825}.
\doiurl{10.1086/339626}.
\adsurl{2002ApJ...570..825B}.
\end{barticle}
\endbibitem

\bibitem[\protect\citeauthoryear{{Caligari}, {Moreno-Insertis}, and
  {Sch{\"u}ssler}}{1995}]{caligari_1995}
\begin{barticle}
\bauthor{\bsnm{{Caligari}}, \binits{P.}},
\bauthor{\bsnm{{Moreno-Insertis}}, \binits{F.}},
\bauthor{\bsnm{{Sch{\"u}ssler}}, \binits{M.}}:
\byear{1995},
\batitle{{Emerging flux tubes in the solar convection zone. 1: Asymmetry, tilt,
  and emergence latitude}}.
\bjtitle{\apj}
\bvolume{441},
\bfpage{886}.
\doiurl{10.1086/175410}.
\adsurl{1995ApJ...441..886C}.
\end{barticle}
\endbibitem

\bibitem[\protect\citeauthoryear{{Caligari}, {Sch{\"u}ssler}, and
  {Moreno-Insertis}}{1998}]{caligari_1998}
\begin{barticle}
\bauthor{\bsnm{{Caligari}}, \binits{P.}},
\bauthor{\bsnm{{Sch{\"u}ssler}}, \binits{M.}},
\bauthor{\bsnm{{Moreno-Insertis}}, \binits{F.}}:
\byear{1998},
\batitle{{Emerging Flux Tubes in the Solar Convection Zone. II. The Influence
  of Initial Conditions}}.
\bjtitle{\apj}
\bvolume{502},
\bfpage{481}.
\doiurl{10.1086/305875}.
\adsurl{1998ApJ...502..481C}.
\end{barticle}
\endbibitem

\bibitem[\protect\citeauthoryear{{Charbonneau}}{2010}]{char_lrsp_2010}
\begin{barticle}
\bauthor{\bsnm{{Charbonneau}}, \binits{P.}}:
\byear{2010},
\batitle{{Dynamo Models of the Solar Cycle}}.
\bjtitle{\lrsp}
\bvolume{7},
\bfpage{3}.
\doiurl{10.12942/lrsp-2010-3}.
\adsurl{2010LRSP....7....3C}.
\end{barticle}
\endbibitem

\bibitem[\protect\citeauthoryear{{Charbonneau}
  \textit{et~al.}}{1999}]{char_1999}
\begin{barticle}
\bauthor{\bsnm{{Charbonneau}}, \binits{P.}},
\bauthor{\bsnm{{Christensen-Dalsgaard}}, \binits{J.}},
\bauthor{\bsnm{{Henning}}, \binits{R.}},
\bauthor{\bsnm{{Larsen}}, \binits{R.M.}},
\bauthor{\bsnm{{Schou}}, \binits{J.}},
\bauthor{\bsnm{{Thompson}}, \binits{M.J.}},
\bauthor{\bsnm{{Tomczyk}}, \binits{S.}}:
\byear{1999},
\batitle{{Helioseismic Constraints on the Structure of the Solar Tachocline}}.
\bjtitle{\apj}
\bvolume{527},
\bfpage{445}.
\doiurl{10.1086/308050}.
\adsurl{1999ApJ...527..445C}.
\end{barticle}
\endbibitem

\bibitem[\protect\citeauthoryear{{Cheng}}{1992}]{cheng_1992}
\begin{barticle}
\bauthor{\bsnm{{Cheng}}, \binits{J.}}:
\byear{1992},
\batitle{{Equations for the motion of an isolated thin magnetic flux tube}}.
\bjtitle{\aap}
\bvolume{264},
\bfpage{243}.
\adsurl{1992A\%26A...264..243C}.
\end{barticle}
\endbibitem

\bibitem[\protect\citeauthoryear{{Choudhuri} and {Gilman}}{1987}]{choud_1987}
\begin{barticle}
\bauthor{\bsnm{{Choudhuri}}, \binits{A.R.}},
\bauthor{\bsnm{{Gilman}}, \binits{P.A.}}:
\byear{1987},
\batitle{{The influence of the Coriolis force on flux tubes rising through the
  solar convection zone}}.
\bjtitle{\apj}
\bvolume{316},
\bfpage{788}.
\doiurl{10.1086/165243}.
\adsurl{1987ApJ...316..788C}.
\end{barticle}
\endbibitem

\bibitem[\protect\citeauthoryear{{Christensen-Dalsgaard}, {Gough}, and
  {Thompson}}{1991}]{jcd_1991}
\begin{barticle}
\bauthor{\bsnm{{Christensen-Dalsgaard}}, \binits{J.}},
\bauthor{\bsnm{{Gough}}, \binits{D.O.}},
\bauthor{\bsnm{{Thompson}}, \binits{M.J.}}:
\byear{1991},
\batitle{{The depth of the solar convection zone}}.
\bjtitle{\apj}
\bvolume{378},
\bfpage{413}.
\doiurl{10.1086/170441}.
\adsurl{1991ApJ...378..413C}.
\end{barticle}
\endbibitem

\bibitem[\protect\citeauthoryear{{Christensen-Dalsgaard}
  \textit{et~al.}}{1996}]{jcd_1996}
\begin{barticle}
\bauthor{\bsnm{{Christensen-Dalsgaard}}, \binits{J.}},
\bauthor{\bsnm{{Dappen}}, \binits{W.}},
\bauthor{\bsnm{{Ajukov}}, \binits{S.V.}},
\bauthor{\bsnm{{Anderson}}, \binits{E.R.}},
\bauthor{\bsnm{{Antia}}, \binits{H.M.}},
\bauthor{\bsnm{{Basu}}, \binits{S.}},
\bauthor{\bsnm{{Baturin}}, \binits{V.A.}},
\bauthor{\bsnm{{Berthomieu}}, \binits{G.}},
\bauthor{\bsnm{{Chaboyer}}, \binits{B.}},
\bauthor{\bsnm{{Chitre}}, \binits{S.M.}},
\bauthor{\bsnm{{Cox}}, \binits{A.N.}},
\bauthor{\bsnm{{Demarque}}, \binits{P.}},
\bauthor{\bsnm{{Donatowicz}}, \binits{J.}},
\bauthor{\bsnm{{Dziembowski}}, \binits{W.A.}},
\bauthor{\bsnm{{Gabriel}}, \binits{M.}},
\bauthor{\bsnm{{Gough}}, \binits{D.O.}},
\bauthor{\bsnm{{Guenther}}, \binits{D.B.}},
\bauthor{\bsnm{{Guzik}}, \binits{J.A.}},
\bauthor{\bsnm{{Harvey}}, \binits{J.W.}},
\bauthor{\bsnm{{Hill}}, \binits{F.}},
\bauthor{\bsnm{{Houdek}}, \binits{G.}},
\bauthor{\bsnm{{Iglesias}}, \binits{C.A.}},
\bauthor{\bsnm{{Kosovichev}}, \binits{A.G.}},
\bauthor{\bsnm{{Leibacher}}, \binits{J.W.}},
\bauthor{\bsnm{{Morel}}, \binits{P.}},
\bauthor{\bsnm{{Proffitt}}, \binits{C.R.}},
\bauthor{\bsnm{{Provost}}, \binits{J.}},
\bauthor{\bsnm{{Reiter}}, \binits{J.}},
\bauthor{\bsnm{{Rhodes}}, \binits{E.J.} \bsuffix{Jr.}},
\bauthor{\bsnm{{Rogers}}, \binits{F.J.}},
\bauthor{\bsnm{{Roxburgh}}, \binits{I.W.}},
\bauthor{\bsnm{{Thompson}}, \binits{M.J.}},
\bauthor{\bsnm{{Ulrich}}, \binits{R.K.}}:
\byear{1996},
\batitle{{The Current State of Solar Modeling}}.
\bjtitle{Science}
\bvolume{272},
\bfpage{1286}.
\doiurl{10.1126/science.272.5266.1286}.
\adsurl{1996Sci...272.1286C}.
\end{barticle}
\endbibitem

\bibitem[\protect\citeauthoryear{{Christensen-Dalsgaard}
  \textit{et~al.}}{2011}]{jcd_2011}
\begin{barticle}
\bauthor{\bsnm{{Christensen-Dalsgaard}}, \binits{J.}},
\bauthor{\bsnm{{Monteiro}}, \binits{M.J.P.F.G.}},
\bauthor{\bsnm{{Rempel}}, \binits{M.}},
\bauthor{\bsnm{{Thompson}}, \binits{M.J.}}:
\byear{2011},
\batitle{{A more realistic representation of overshoot at the base of the solar
  convective envelope as seen by helioseismology}}.
\bjtitle{\mnras}
\bvolume{414},
\bfpage{1158}.
\doiurl{10.1111/j.1365-2966.2011.18460.x}.
\adsurl{2011MNRAS.414.1158C}.
\end{barticle}
\endbibitem

\bibitem[\protect\citeauthoryear{{Dasi-Espuig}
  \textit{et~al.}}{2010}]{dasi_2010}
\begin{barticle}
\bauthor{\bsnm{{Dasi-Espuig}}, \binits{M.}},
\bauthor{\bsnm{{Solanki}}, \binits{S.K.}},
\bauthor{\bsnm{{Krivova}}, \binits{N.A.}},
\bauthor{\bsnm{{Cameron}}, \binits{R.}},
\bauthor{\bsnm{{Pe{\~n}uela}}, \binits{T.}}:
\byear{2010},
\batitle{{Sunspot group tilt angles and the strength of the solar cycle}}.
\bjtitle{\aap}
\bvolume{518},
\bfpage{A7}.
\doiurl{10.1051/0004-6361/201014301}.
\adsurl{http://cdsads.u-strasbg.fr/abs/2010A\%26A...518A...7D}.
\end{barticle}
\endbibitem

\bibitem[\protect\citeauthoryear{{Defouw}}{1976}]{defouw_1976}
\begin{barticle}
\bauthor{\bsnm{{Defouw}}, \binits{R.J.}}:
\byear{1976},
\batitle{{Wave propagation along a magnetic tube}}.
\bjtitle{\apj}
\bvolume{209},
\bfpage{266}.
\doiurl{10.1086/154717}.
\adsurl{1976ApJ...209..266D}.
\end{barticle}
\endbibitem

\bibitem[\protect\citeauthoryear{{D'Silva} and {Choudhuri}}{1993}]{dsilva_1993}
\begin{barticle}
\bauthor{\bsnm{{D'Silva}}, \binits{S.}},
\bauthor{\bsnm{{Choudhuri}}, \binits{A.R.}}:
\byear{1993},
\batitle{{A theoretical model for tilts of bipolar magnetic regions}}.
\bjtitle{\aap}
\bvolume{272},
\bfpage{621}.
\adsurl{1993A\%26A...272..621D}.
\end{barticle}
\endbibitem

\bibitem[\protect\citeauthoryear{{Fan}}{2009}]{fan_2009}
\begin{barticle}
\bauthor{\bsnm{{Fan}}, \binits{Y.}}:
\byear{2009},
\batitle{{Magnetic Fields in the Solar Convection Zone}}.
\bjtitle{\lrsp}
\bvolume{6},
\bfpage{4}.
\doiurl{10.12942/lrsp-2009-4}.
\adsurl{2009LRSP....6....4F}.
\end{barticle}
\endbibitem

\bibitem[\protect\citeauthoryear{{Fan} and {Fisher}}{1996}]{fan_1996}
\begin{barticle}
\bauthor{\bsnm{{Fan}}, \binits{Y.}},
\bauthor{\bsnm{{Fisher}}, \binits{G.H.}}:
\byear{1996},
\batitle{{Radiative Heating and the Buoyant Rise of Magnetic Flux Tubes in the
  Solar interior}}.
\bjtitle{\solphys}
\bvolume{166},
\bfpage{17}.
\doiurl{10.1007/BF00179354}.
\adsurl{1996SoPh..166...17F}.
\end{barticle}
\endbibitem

\bibitem[\protect\citeauthoryear{{Fan}, {Fisher}, and
  {Deluca}}{1993}]{fan_1993}
\begin{barticle}
\bauthor{\bsnm{{Fan}}, \binits{Y.}},
\bauthor{\bsnm{{Fisher}}, \binits{G.H.}},
\bauthor{\bsnm{{Deluca}}, \binits{E.E.}}:
\byear{1993},
\batitle{{The origin of morphological asymmetries in bipolar active regions}}.
\bjtitle{\apj}
\bvolume{405},
\bfpage{390}.
\doiurl{10.1086/172370}.
\adsurl{1993ApJ...405..390F}.
\end{barticle}
\endbibitem

\bibitem[\protect\citeauthoryear{{Fan}, {Fisher}, and
  {McClymont}}{1994}]{fan_1994}
\begin{barticle}
\bauthor{\bsnm{{Fan}}, \binits{Y.}},
\bauthor{\bsnm{{Fisher}}, \binits{G.H.}},
\bauthor{\bsnm{{McClymont}}, \binits{A.N.}}:
\byear{1994},
\batitle{{Dynamics of emerging active region flux loops}}.
\bjtitle{\apj}
\bvolume{436},
\bfpage{907}.
\doiurl{10.1086/174967}.
\adsurl{http://cdsads.u-strasbg.fr/abs/1994ApJ...436..907F}.
\end{barticle}
\endbibitem

\bibitem[\protect\citeauthoryear{{Ferriz-Mas} and
  {Sch{\"u}ssler}}{1993}]{ferriz_1993}
\begin{barticle}
\bauthor{\bsnm{{Ferriz-Mas}}, \binits{A.}},
\bauthor{\bsnm{{Sch{\"u}ssler}}, \binits{M.}}:
\byear{1993},
\batitle{{Instabilities of magnetic flux tubes in a stellar convection zone I.
  Equatorial flux rings in differentially rotating stars}}.
\bjtitle{\gafd}
\bvolume{72},
\bfpage{209}.
\doiurl{10.1080/03091929308203613}.
\adsurl{1993GApFD..72..209F}.
\end{barticle}
\endbibitem

\bibitem[\protect\citeauthoryear{{Ferriz-Mas}, {Sch{\"u}ssler}, and
  {Anton}}{1989}]{ferriz_1989}
\begin{barticle}
\bauthor{\bsnm{{Ferriz-Mas}}, \binits{A.}},
\bauthor{\bsnm{{Sch{\"u}ssler}}, \binits{M.}},
\bauthor{\bsnm{{Anton}}, \binits{V.}}:
\byear{1989},
\batitle{{Dynamics of magnetic flux concentrations - The second-order thin flux
  tube approximation}}.
\bjtitle{\aap}
\bvolume{210},
\bfpage{425}.
\adsurl{1989A\%26A...210..425F}.
\end{barticle}
\endbibitem

\bibitem[\protect\citeauthoryear{{Fisher}, {Fan}, and
  {Howard}}{1995}]{fisher_1995}
\begin{barticle}
\bauthor{\bsnm{{Fisher}}, \binits{G.H.}},
\bauthor{\bsnm{{Fan}}, \binits{Y.}},
\bauthor{\bsnm{{Howard}}, \binits{R.F.}}:
\byear{1995},
\batitle{{Comparisons between theory and observation of active region tilts}}.
\bjtitle{\apj}
\bvolume{438},
\bfpage{463}.
\doiurl{10.1086/175090}.
\adsurl{http://cdsads.u-strasbg.fr/abs/1995ApJ...438..463F}.
\end{barticle}
\endbibitem

\bibitem[\protect\citeauthoryear{{Galloway} and {Weiss}}{1981}]{galloway_1981}
\begin{barticle}
\bauthor{\bsnm{{Galloway}}, \binits{D.J.}},
\bauthor{\bsnm{{Weiss}}, \binits{N.O.}}:
\byear{1981},
\batitle{{Convection and magnetic fields in stars}}.
\bjtitle{\apj}
\bvolume{243},
\bfpage{945}.
\doiurl{10.1086/158659}.
\adsurl{1981ApJ...243..945G}.
\end{barticle}
\endbibitem

\bibitem[\protect\citeauthoryear{{Gilman}}{2000}]{gilman_2000}
\begin{barticle}
\bauthor{\bsnm{{Gilman}}, \binits{P.A.}}:
\byear{2000},
\batitle{{Fluid Dynamics and MHD of the Solar Convection Zone and Tachocline:
  Current Understanding and Unsolved Problems - (Invited Review)}}.
\bjtitle{\solphys}
\bvolume{192},
\bfpage{27}.
\adsurl{2000SoPh..192...27G}.
\end{barticle}
\endbibitem

\bibitem[\protect\citeauthoryear{{Hale} \textit{et~al.}}{1919}]{hale_1919}
\begin{barticle}
\bauthor{\bsnm{{Hale}}, \binits{G.E.}},
\bauthor{\bsnm{{Ellerman}}, \binits{F.}},
\bauthor{\bsnm{{Nicholson}}, \binits{S.B.}},
\bauthor{\bsnm{{Joy}}, \binits{A.H.}}:
\byear{1919},
\batitle{{The Magnetic Polarity of Sun-Spots}}.
\bjtitle{\apj}
\bvolume{49},
\bfpage{153}.
\doiurl{10.1086/142452}.
\adsurl{1919ApJ....49..153H}.
\end{barticle}
\endbibitem

\bibitem[\protect\citeauthoryear{{Howard}}{1996}]{howard_1996}
\begin{barticle}
\bauthor{\bsnm{{Howard}}, \binits{R.F.}}:
\byear{1996},
\batitle{{Axial Tilt Angles of Active Regions}}.
\bjtitle{\solphys}
\bvolume{169},
\bfpage{293}.
\doiurl{10.1007/BF00190606}.
\adsurl{http://cdsads.u-strasbg.fr/abs/1996SoPh..169..293H}.
\end{barticle}
\endbibitem

\bibitem[\protect\citeauthoryear{{Li} and {Ulrich}}{2012}]{li_2012}
\begin{barticle}
\bauthor{\bsnm{{Li}}, \binits{J.}},
\bauthor{\bsnm{{Ulrich}}, \binits{R.K.}}:
\byear{2012},
\batitle{{Long-term Measurements of Sunspot Magnetic Tilt Angles}}.
\bjtitle{\apj}
\bvolume{758},
\bfpage{115}.
\doiurl{10.1088/0004-637X/758/2/115}.
\adsurl{2012ApJ...758..115L}.
\end{barticle}
\endbibitem

\bibitem[\protect\citeauthoryear{{McClintock} and
  {Norton}}{2013}]{mcclintock_2013}
\begin{barticle}
\bauthor{\bsnm{{McClintock}}, \binits{B.H.}},
\bauthor{\bsnm{{Norton}}, \binits{A.A.}}:
\byear{2013},
\batitle{{Recovering Joy's Law as a Function of Solar Cycle, Hemisphere, and
  Longitude}}.
\bjtitle{\solphys}
\bvolume{287},
\bfpage{215}.
\doiurl{10.1007/s11207-013-0338-0}.
\adsurl{http://cdsads.u-strasbg.fr/abs/2013SoPh..287..215M}.
\end{barticle}
\endbibitem

\bibitem[\protect\citeauthoryear{{Miesch}}{2005}]{miesch_lrsp}
\begin{barticle}
\bauthor{\bsnm{{Miesch}}, \binits{M.S.}}:
\byear{2005},
\batitle{{Large-Scale Dynamics of the Convection Zone and Tachocline}}.
\bjtitle{\lrsp}
\bvolume{2},
\bfpage{1}.
\doiurl{10.12942/lrsp-2005-1}.
\adsurl{2005LRSP....2....1M}.
\end{barticle}
\endbibitem

\bibitem[\protect\citeauthoryear{{Miesch} and {Toomre}}{2009}]{miesch_2009}
\begin{barticle}
\bauthor{\bsnm{{Miesch}}, \binits{M.S.}},
\bauthor{\bsnm{{Toomre}}, \binits{J.}}:
\byear{2009},
\batitle{{Turbulence, Magnetism, and Shear in Stellar Interiors}}.
\bjtitle{Annual Review of Fluid Mechanics}
\bvolume{41},
\bfpage{317}.
\doiurl{10.1146/annurev.fluid.010908.165215}.
\adsurl{2009AnRFM..41..317M}.
\end{barticle}
\endbibitem

\bibitem[\protect\citeauthoryear{{Miesch}, {Brun}, and
  {Toomre}}{2006}]{miesch_apj_2006}
\begin{barticle}
\bauthor{\bsnm{{Miesch}}, \binits{M.S.}},
\bauthor{\bsnm{{Brun}}, \binits{A.S.}},
\bauthor{\bsnm{{Toomre}}, \binits{J.}}:
\byear{2006},
\batitle{{Solar Differential Rotation Influenced by Latitudinal Entropy
  Variations in the Tachocline}}.
\bjtitle{\apj}
\bvolume{641},
\bfpage{618}.
\doiurl{10.1086/499621}.
\adsurl{2006ApJ...641..618M}.
\end{barticle}
\endbibitem

\bibitem[\protect\citeauthoryear{{Monteiro}, {Christensen-Dalsgaard}, and
  {Thompson}}{1994}]{monteiro_1994}
\begin{barticle}
\bauthor{\bsnm{{Monteiro}}, \binits{M.J.P.F.G.}},
\bauthor{\bsnm{{Christensen-Dalsgaard}}, \binits{J.}},
\bauthor{\bsnm{{Thompson}}, \binits{M.J.}}:
\byear{1994},
\batitle{{Seismic study of overshoot at the base of the solar convective
  envelope}}.
\bjtitle{\aap}
\bvolume{283},
\bfpage{247}.
\adsurl{1994A\%26A...283..247M}.
\end{barticle}
\endbibitem

\bibitem[\protect\citeauthoryear{{Moreno-Insertis}}{1986}]{moreno_1986}
\begin{barticle}
\bauthor{\bsnm{{Moreno-Insertis}}, \binits{F.}}:
\byear{1986},
\batitle{{Nonlinear time-evolution of kink-unstable magnetic flux tubes in the
  convective zone of the sun}}.
\bjtitle{\aap}
\bvolume{166},
\bfpage{291}.
\adsurl{1986A\%26A...166..291M}.
\end{barticle}
\endbibitem

\bibitem[\protect\citeauthoryear{{Moreno-Insertis} and
  {Emonet}}{1996}]{moreno_1996}
\begin{barticle}
\bauthor{\bsnm{{Moreno-Insertis}}, \binits{F.}},
\bauthor{\bsnm{{Emonet}}, \binits{T.}}:
\byear{1996},
\batitle{{The Rise of Twisted Magnetic Tubes in a Stratified Medium}}.
\bjtitle{\apjl}
\bvolume{472},
\bfpage{L53}.
\doiurl{10.1086/310360}.
\adsurl{1996ApJ...472L..53M}.
\end{barticle}
\endbibitem

\bibitem[\protect\citeauthoryear{{Moreno-Insertis}, {Sch\"ussler}, and
  {Ferriz-Mas}}{1992}]{moreno_1992}
\begin{barticle}
\bauthor{\bsnm{{Moreno-Insertis}}, \binits{F.}},
\bauthor{\bsnm{{Sch\"ussler}}, \binits{M.}},
\bauthor{\bsnm{{Ferriz-Mas}}, \binits{A.}}:
\byear{1992},
\batitle{{Storage of magnetic flux tubes in a convective overshoot region}}.
\bjtitle{\aap}
\bvolume{264},
\bfpage{686}.
\adsurl{1992A\%26A...264..686M}.
\end{barticle}
\endbibitem

\bibitem[\protect\citeauthoryear{{Moreno-Insertis}, {Sch{\"u}ssler}, and
  {Glampedakis}}{2002}]{moreno_2002}
\begin{barticle}
\bauthor{\bsnm{{Moreno-Insertis}}, \binits{F.}},
\bauthor{\bsnm{{Sch{\"u}ssler}}, \binits{M.}},
\bauthor{\bsnm{{Glampedakis}}, \binits{K.}}:
\byear{2002},
\batitle{{Thermal properties of magnetic flux tubes. I. Solution of the
  diffusion problem}}.
\bjtitle{\aap}
\bvolume{388},
\bfpage{1022}.
\doiurl{10.1051/0004-6361:20020488}.
\adsurl{2002A\%26A...388.1022M}.
\end{barticle}
\endbibitem

\bibitem[\protect\citeauthoryear{{Parker}}{1979}]{parker_book}
\begin{bbook}
\bauthor{\bsnm{{Parker}}, \binits{E.N.}}:
\byear{1979},
\bbtitle{{Cosmical magnetic fields: Their origin and their activity}}.
\adsurl{1979cmft.book.....P}.
\end{bbook}
\endbibitem

\bibitem[\protect\citeauthoryear{{Pidatella} and {Stix}}{1986}]{pidatella_1986}
\begin{barticle}
\bauthor{\bsnm{{Pidatella}}, \binits{R.M.}},
\bauthor{\bsnm{{Stix}}, \binits{M.}}:
\byear{1986},
\batitle{{Convective overshoot at the base of the sun's convection zone}}.
\bjtitle{\aap}
\bvolume{157},
\bfpage{338}.
\adsurl{1986A\%26A...157..338P}.
\end{barticle}
\endbibitem

\bibitem[\protect\citeauthoryear{{Rempel}}{2003}]{rempel_2003}
\begin{barticle}
\bauthor{\bsnm{{Rempel}}, \binits{M.}}:
\byear{2003},
\batitle{{Thermal properties of magnetic flux tubes. II. Storage of flux in the
  solar overshoot region}}.
\bjtitle{\aap}
\bvolume{397},
\bfpage{1097}.
\doiurl{10.1051/0004-6361:20021594}.
\adsurl{2003A\%26A...397.1097R}.
\end{barticle}
\endbibitem

\bibitem[\protect\citeauthoryear{{Rempel}}{2004}]{rempel_2004}
\begin{barticle}
\bauthor{\bsnm{{Rempel}}, \binits{M.}}:
\byear{2004},
\batitle{{Overshoot at the Base of the Solar Convection Zone: A Semianalytical
  Approach}}.
\bjtitle{\apj}
\bvolume{607},
\bfpage{1046}.
\doiurl{10.1086/383605}.
\adsurl{2004ApJ...607.1046R}.
\end{barticle}
\endbibitem

\bibitem[\protect\citeauthoryear{{Roberts} and {Webb}}{1978}]{roberts_1978}
\begin{barticle}
\bauthor{\bsnm{{Roberts}}, \binits{B.}},
\bauthor{\bsnm{{Webb}}, \binits{A.R.}}:
\byear{1978},
\batitle{{Vertical motions in an intense magnetic flux tube}}.
\bjtitle{\solphys}
\bvolume{56},
\bfpage{5}.
\doiurl{10.1007/BF00152630}.
\adsurl{1978SoPh...56....5R}.
\end{barticle}
\endbibitem

\bibitem[\protect\citeauthoryear{{Schmitt}, {Rosner}, and
  {Bohn}}{1984}]{schmitt_1984}
\begin{barticle}
\bauthor{\bsnm{{Schmitt}}, \binits{J.H.M.M.}},
\bauthor{\bsnm{{Rosner}}, \binits{R.}},
\bauthor{\bsnm{{Bohn}}, \binits{H.U.}}:
\byear{1984},
\batitle{{The overshoot region at the bottom of the solar convection zone}}.
\bjtitle{\apj}
\bvolume{282},
\bfpage{316}.
\doiurl{10.1086/162205}.
\adsurl{1984ApJ...282..316S}.
\end{barticle}
\endbibitem

\bibitem[\protect\citeauthoryear{{Skaley} and {Stix}}{1991}]{skaley_1991}
\begin{barticle}
\bauthor{\bsnm{{Skaley}}, \binits{D.}},
\bauthor{\bsnm{{Stix}}, \binits{M.}}:
\byear{1991},
\batitle{{The overshoot layer at the base of the solar convection zone}}.
\bjtitle{\aap}
\bvolume{241},
\bfpage{227}.
\adsurl{1991A\%26A...241..227S}.
\end{barticle}
\endbibitem

\bibitem[\protect\citeauthoryear{{Spiegel} and {Weiss}}{1980}]{spiegel_1980}
\begin{barticle}
\bauthor{\bsnm{{Spiegel}}, \binits{E.A.}},
\bauthor{\bsnm{{Weiss}}, \binits{N.O.}}:
\byear{1980},
\batitle{{Magnetic activity and variations in solar luminosity}}.
\bjtitle{\nat}
\bvolume{287},
\bfpage{616}.
\doiurl{10.1038/287616a0}.
\adsurl{1980Natur.287..616S}.
\end{barticle}
\endbibitem

\bibitem[\protect\citeauthoryear{{Spruit}}{1981a}]{spruit_1981a}
\begin{barticle}
\bauthor{\bsnm{{Spruit}}, \binits{H.C.}}:
\byear{1981}a,
\batitle{{Equations for thin flux tubes in ideal MHD}}.
\bjtitle{\aap}
\bvolume{102},
\bfpage{129}.
\adsurl{1981A\%26A...102..129S}.
\end{barticle}
\endbibitem

\bibitem[\protect\citeauthoryear{{Spruit}}{1981b}]{spruit_1981b}
\begin{barticle}
\bauthor{\bsnm{{Spruit}}, \binits{H.C.}}:
\byear{1981}b,
\batitle{{Motion of magnetic flux tubes in the solar convection zone and
  chromosphere}}.
\bjtitle{\aap}
\bvolume{98},
\bfpage{155}.
\adsurl{1981A\%26A....98..155S}.
\end{barticle}
\endbibitem

\bibitem[\protect\citeauthoryear{{Stenflo} and
  {Kosovichev}}{2012}]{stenflo_2012}
\begin{barticle}
\bauthor{\bsnm{{Stenflo}}, \binits{J.O.}},
\bauthor{\bsnm{{Kosovichev}}, \binits{A.G.}}:
\byear{2012},
\batitle{{Bipolar Magnetic Regions on the Sun: Global Analysis of the SOHO/MDI
  Data Set}}.
\bjtitle{\apj}
\bvolume{745},
\bfpage{129}.
\doiurl{10.1088/0004-637X/745/2/129}.
\adsurl{http://cdsads.u-strasbg.fr/abs/2012ApJ...745..129S}.
\end{barticle}
\endbibitem

\bibitem[\protect\citeauthoryear{{Thompson}
  \textit{et~al.}}{2003}]{thompson_araa_2003}
\begin{barticle}
\bauthor{\bsnm{{Thompson}}, \binits{M.J.}},
\bauthor{\bsnm{{Christensen-Dalsgaard}}, \binits{J.}},
\bauthor{\bsnm{{Miesch}}, \binits{M.S.}},
\bauthor{\bsnm{{Toomre}}, \binits{J.}}:
\byear{2003},
\batitle{{The Internal Rotation of the Sun}}.
\bjtitle{\araa}
\bvolume{41},
\bfpage{599}.
\doiurl{10.1146/annurev.astro.41.011802.094848}.
\adsurl{2003ARA\%26A..41..599T}.
\end{barticle}
\endbibitem

\bibitem[\protect\citeauthoryear{{Tobias} \textit{et~al.}}{2001}]{tobias_2001}
\begin{barticle}
\bauthor{\bsnm{{Tobias}}, \binits{S.M.}},
\bauthor{\bsnm{{Brummell}}, \binits{N.H.}},
\bauthor{\bsnm{{Clune}}, \binits{T.L.}},
\bauthor{\bsnm{{Toomre}}, \binits{J.}}:
\byear{2001},
\batitle{Transport and storage of magnetic field by overshooting turbulent
  compressible convection}.
\bjtitle{\apj}
\bvolume{549},
\bfpage{1183}.
\doiurl{10.1086/319448}.
\adsurl{2001ApJ...549.1183T}.
\end{barticle}
\endbibitem

\bibitem[\protect\citeauthoryear{{van Ballegooijen}}{1982}]{vanballe_1982}
\begin{barticle}
\bauthor{\bsnm{{van Ballegooijen}}, \binits{A.A.}}:
\byear{1982},
\batitle{{The overshoot layer at the base of the solar convective zone and the
  problem of magnetic flux storage}}.
\bjtitle{\aap}
\bvolume{113},
\bfpage{99}.
\adsurl{1982A\%26A...113...99V}.
\end{barticle}
\endbibitem

\bibitem[\protect\citeauthoryear{{van Ballegooijen} and
  {Choudhuri}}{1988}]{vanballe_1988}
\begin{barticle}
\bauthor{\bsnm{{van Ballegooijen}}, \binits{A.A.}},
\bauthor{\bsnm{{Choudhuri}}, \binits{A.R.}}:
\byear{1988},
\batitle{{The possible role of meridional flows in suppressing magnetic
  buoyancy}}.
\bjtitle{\apj}
\bvolume{333},
\bfpage{965}.
\doiurl{10.1086/166805}.
\adsurl{1988ApJ...333..965V}.
\end{barticle}
\endbibitem

\bibitem[\protect\citeauthoryear{{Wang} and {Zirin}}{1989}]{wang_1989}
\begin{barticle}
\bauthor{\bsnm{{Wang}}, \binits{H.}},
\bauthor{\bsnm{{Zirin}}, \binits{H.}}:
\byear{1989},
\batitle{{Study of supergranules}}.
\bjtitle{\solphys}
\bvolume{120},
\bfpage{1}.
\doiurl{10.1007/BF00148532}.
\adsurl{1989SoPh..120....1W}.
\end{barticle}
\endbibitem

\bibitem[\protect\citeauthoryear{{Wang} and
  {Sheeley}}{1989}]{wang_sheeley_1989}
\begin{barticle}
\bauthor{\bsnm{{Wang}}, \binits{Y.-M.}},
\bauthor{\bsnm{{Sheeley}}, \binits{N.R.} \bsuffix{Jr.}}:
\byear{1989},
\batitle{{Average properties of bipolar magnetic regions during sunspot cycle
  21}}.
\bjtitle{\solphys}
\bvolume{124},
\bfpage{81}.
\doiurl{10.1007/BF00146521}.
\adsurl{http://cdsads.u-strasbg.fr/abs/1989SoPh..124...81W}.
\end{barticle}
\endbibitem

\bibitem[\protect\citeauthoryear{{Weber}, {Fan}, and
  {Miesch}}{2011}]{weber_2011}
\begin{barticle}
\bauthor{\bsnm{{Weber}}, \binits{M.A.}},
\bauthor{\bsnm{{Fan}}, \binits{Y.}},
\bauthor{\bsnm{{Miesch}}, \binits{M.S.}}:
\byear{2011},
\batitle{{The Rise of Active Region Flux Tubes in the Turbulent Solar
  Convective Envelope}}.
\bjtitle{\apj}
\bvolume{741},
\bfpage{11}.
\doiurl{10.1088/0004-637X/741/1/11}.
\adsurl{2011ApJ...741...11W}.
\end{barticle}
\endbibitem

\bibitem[\protect\citeauthoryear{{Weber}, {Fan}, and
  {Miesch}}{2013}]{weber_solphys_2013}
\begin{barticle}
\bauthor{\bsnm{{Weber}}, \binits{M.A.}},
\bauthor{\bsnm{{Fan}}, \binits{Y.}},
\bauthor{\bsnm{{Miesch}}, \binits{M.S.}}:
\byear{2013},
\batitle{{Comparing Simulations of Rising Flux Tubes Through the Solar
  Convection Zone with Observations of Solar Active Regions: Constraining the
  Dynamo Field Strength}}.
\bjtitle{\solphys}
\bvolume{287},
\bfpage{239}.
\doiurl{10.1007/s11207-012-0093-7}.
\adsurl{2013SoPh..287..239W}.
\end{barticle}
\endbibitem

\bibitem[\protect\citeauthoryear{{Zahn}}{1991}]{zahn_1991}
\begin{barticle}
\bauthor{\bsnm{{Zahn}}, \binits{J.-P.}}:
\byear{1991},
\batitle{{Convective penetration in stellar interiors}}.
\bjtitle{\aap}
\bvolume{252},
\bfpage{179}.
\adsurl{1991A\%26A...252..179Z}.
\end{barticle}
\endbibitem

\end{thebibliography}
%
%
%
%

\end{article} 
\end{document}